\pdfoutput=1

\documentclass[twocolumn,draft,prd,superscriptaddress,showpacs,floatfix%
,preprintnumbers,nofootinbib]{revtex4}

\usepackage[final]{epsfig}
\usepackage{bm}
\usepackage[usenames]{color}

\begin{document}

\title{Neutrino emission characteristics and detection opportunities
based on three-dimensional supernova simulations}

\author{Irene Tamborra}
\affiliation{GRAPPA Institute, University of
Amsterdam,
Science Park 904, 1098 XH Amsterdam, The Netherlands}

\author{Georg Raffelt}
\affiliation{Max-Planck-Institut f\"ur Physik (Werner-Heisenberg-Institut),
F\"ohringer Ring 6, 80805 M\"unchen, Germany}

\author{Florian Hanke}
\affiliation{Max-Planck-Institut f\"ur Astrophysik,
Karl-Schwarzschild-Str.~1, 85748 Garching, Germany}
\affiliation{Physik Department, Technische Universit\"at M\"unchen, James-Franck-Str.~1, 85748 Garching, Germany}

\author{Hans-Thomas Janka}
\affiliation{Max-Planck-Institut f\"ur Astrophysik,
Karl-Schwarzschild-Str.~1, 85748 Garching, Germany}

\author{Bernhard M\"uller}
\affiliation{Monash Center for Astrophysics, School of
  Mathematical Sciences, Building 28, Monash University, Victoria
  3800, Australia}

\date{\today}

\preprint{MPP-2014-76}

\begin{abstract}
The neutrino emission characteristics of the first full-scale three-dimensional
supernova simulations with sophisticated three-flavor neutrino transport for three
models with masses 11.2, 20 and $27\,M_\odot$ are evaluated in detail.
All the studied progenitors  show the  expected hydrodynamical instabilities in the form 
of large-scale convective overturn. In addition, the recently identified LESA phenomenon
(lepton-number emission self-sustained asymmetry) is generic for all our cases. Pronounced 
SASI (standing accretion-shock instability) activity appears in the  20 and $27\,M_\odot$ cases, 
partly in the form of a spiral mode, inducing large but direction and flavor-dependent modulations 
of neutrino emission. These modulations can be clearly identified in the existing IceCube and future 
Hyper-Kamiokande detectors, depending on distance and detector location relative to the main SASI sloshing
direction. 
\end{abstract}


\maketitle

\section{Introduction}
\label{sec:intro}

The neutrino signal of the next nearby core-collapse supernova (SN) will be
measured in many detectors that will register tens to hundreds of events,
assuming a fiducial distance in the galaxy of 10~kpc~\cite{Scholberg:2012id}.
The largest statistics will be provided by
Super-Kamiokande~\cite{Abe:2013gga, Mori:2013wua} with roughly $10^4$ and
IceCube \cite{Abbasi:2011ss, Demiroers:2011am, Aartsen:2013nla} with roughly
$10^6$ events, the latter without event-by-event energy information. In the
context of neutrino oscillation physics, additional large detectors are in
different phases of planning, notably JUNO~\cite{Li:2014qca}, a 20~kt liquid
scintillator detector, Hyper-Kamiokande~\cite{Abe:2011ts}, a megaton water
Cherenkov detector, and a 30~kt liquid-argon time-projection chamber
\cite{Agarwalla:2013kaa, LBNE}. The main problem, of course, is that galactic
SNe are rare, perhaps one every few decades~\cite{vandenBergh:1994,
Li:2010kd, Reed:2005en, FaucherGiguere:2005ny, Keane:2008jj, Diehl:2006cf,
Strom:1994, Tammann:1994ev, Adams:2013ana, Alekseev:1993dy}. Clearly we
should prepare well for such a once-in-a-lifetime opportunity and should
understand in advance what could be learnt from such an observation.

The low-statistics neutrino signal of SN~1987A has confirmed the general
picture of stellar core collapse, but was too sparse to extract much
astrophysical detail~\cite{Schramm:1987ra}. On the other hand, it has
provided many useful particle-physics lessons, notably on the possible
energy loss in new forms of radiation such as axions \cite{Raffelt:1990yz,
Raffelt:1999tx}. A future observation will refine
such arguments, but the real benefit of high statistics may be
detailed astrophysical information on the physics of core
collapse~\cite{Totani:1997vj, Lund:2010kh, Lund:2012vm, Brandt:2010xa, Tamborra:2013laa, OConnor:2013, Mueller:2014rna}.
Thirty years after the formulation of the neutrino-driven delayed-explosion
paradigm by Bethe and Wilson~\cite{Bethe:1984ux, Bethe:1990mw}, we still can
not be sure that their theory is not missing some important
ingredient~\cite{Janka:2012wk}.

According to their scenario, a shock wave forms at core bounce, stalls after
reaching a radius of 100--200~km, and is revived by neutrino heating after
tens to hundreds of milliseconds, depending on progenitor properties and
accretion rate of stellar matter that continues to collapse. Moreover,
observed SN asymmetries and two-dimensional (2D)~\cite{Herant:1994dd,Burrows:1995ww,
Janka:1996,Marek:2008qi,Brandt:2010xa, Marek:2007gr, Mueller:2012is, Mueller:2012ak, Murphy:2008dw, Nordhaus:2010uk}
and 3D~\cite{Fryer:2002,Fryer:2003jj,Iwakami:2007ie, Wongwathanarat:2010ip,
Muller:2011yi, Hanke:2011jf, Takiwaki:2011db, Burrows:2012yk, Ott:2013}
hydrodynamical simulations imply that SN explosions are inherently
multi-dimensional. During the accretion phase, large-scale convective
overturn develops in the neutrino-heated postshock layer and the standing
accretion shock instability (SASI)~\cite{Blondin:2002sm}
can arise, involving global deformation
and sloshing motions of the shock front~\cite{Blondin:2002sm, Scheck:2007gw,
Foglizzo:2011aa}.

In the course of the present research project we have recently discovered the
LESA phenomenon (``lepton-number emission self-sustained asymmetry'')
\cite{Tamborra:2014aua}. The deleptonization ($\nu_e$ minus $\bar\nu_e$) flux
during the accretion phase develops a pronounced dipole pattern, i.e. the
lepton-number flux emerges predominantly in one hemisphere. We have
identified a feed-back loop as the likely cause of this effect. Its elements
are asymmetric accretion caused by shock-wave deformation and asymmetric
neutrino heating behind the shock front causing the shock-front deformation.
It is not yet clear if LESA is a benign curiosity of multi-dimensional SN
physics or an important player in the overall core-collapse phenomenology,
perhaps in conjunction with neutrino flavor conversion. Either way, its
discovery certainly shows that in multi-dimensional SN models there is room
for hitherto unsuspected new phenomena.

The various hydrodynamical instabilities appearing in 3D core collapse during
the phase of a standing accretion shock imply that the neutrino signal expected
from the next nearby SN can show fast modulations and depends on observer
location relative to the main direction of SASI sloshing and relative to the
LESA dipole direction. The main purpose of our paper is to explore these
issues based on our current portfolio of 3D core-collapse models with
full-scale three-flavor neutrino transport. The progenitor masses are 11.2,
20 and $27\,M_\odot$, all of them show the LESA phenomenon and the two
heavier models show pronounced SASI activity.

The present paper expands on our earlier {\em Physical Review
Letter\/}~\cite{Tamborra:2013laa} where we have reported the appearance of
signal modulations by SASI that are detectable in IceCube and the future
Hyper-Kamiokande detector. In the context of 2D models, this point had been
made earlier~\cite{Lund:2010kh}. On the other hand, it had also been shown
that convective overturn alone produces signal modulations that can be
detected only if the SN is very close~\cite{Lund:2012vm}. Therefore,
detectable signal modulations are typically tied to the appearance of SASI.

A vigorous debate among SN modelers had revolved around the question if SASI
indeed appears in 3D models or if its growth would be suppressed by
large-scale convective overturn~\cite{Burrows:2012yk, Murphy:2012id,
Dolence:2012kh}. Meanwhile SASI activity in 3D SN models with
different neutrino treatments was found by several
authors~\cite{Ott:2013,Hanke:2013ena,Couch:2013kma},
but such a convergence of qualitative numerical conclusions, of
course, leaves open the question of what actually happens in nature. The
appearance of the SASI is driven by progenitor-dependent conditions, which
determine the growth rates of the SASI and convective instability in the
postshock accretion layer~\cite{Scheck:2007gw}.
A neutrino observation of SASI modulations would
be a unique smoking gun to prove its very existence in real-life
core-collapse events.

When we studied more closely how the neutrino emission characteristics depend
on observer direction we noticed a pronounced asymmetry in the lepton-number
flux, whereas the overall neutrino luminosity is nearly spherically symmetric
except for the SASI modulations~\cite{Tamborra:2014aua}.
A detailed study of
the various elements of this puzzling LESA phenomenon, however, has yielded
support for its possibly physical origin. In
particular, we believe that we have identified the feedback loop driving this
new neutrino-hydrodynamical instability. Nevertheless, we cannot exclude
that LESA is a numerical artifact and the final verdict
depends on LESA being reproduced 
by 3D models with true multi-D transport (for discussions of multi-D transport effects in 3D, see 
Ref.~\cite{Sumiyoshi:2014qua}, and in 2D, see Ref.~\cite{Dolence:2014rwa}).

A directional dependence of this sort is not immediately obvious in the usual
visualization of multi-D hydrodynamical simulations. Extracting the neutrino
signal characteristics as a function of observer direction requires a
significant amount of dedicated post-processing. In this sense, our study is
also meant to encourage other SN modelers to show this sort of information
which is important for neutrino signal detection and studies of flavor
conversion. Our procedure for an efficient extraction of this
information may be useful for other authors as well.

Of course, our discussion pertains exclusively to the SN accretion phase
where hydrodynamical instabilities are a key element and which must
ultimately lead to the explosion. For the initial collapse and bounce
phase, perhaps up to about 100~ms after bounce, spherical symmetry remains
a good approximation in discussing the neutrino emission. The neutrino
signal during this early phase is surprisingly independent of model
details~\cite{Kachelriess:2004ds, Serpico:2011ir}. Likewise, after the
explosion has taken off, the subsequent phase of proto-neutron star cooling
is again governed by spherically symmetric emission.
These three phases should be seen as
distinct episodes, testing very different aspects of hydrodynamics as well
as nuclear and particle physics.

Our paper begins in Sec.~II with a summary of the main features of our SN
models. In Sec.~III we discuss the features of the neutrino signal from our
three SN progenitors and the features of the LESA phenomenon
in the presence of SASI. In Sec.~IV we review the role of neutrino oscillations,
while in Sec.~V we focus on the detection perspectives of the signal
modulation in IceCube and Hyper-Kamiokande. Discussion and conclusions will
be presented in Sec.~VI.

\section{Numerical Supernova Models}
\label{sec:models}

Our SN simulations were performed with the neutrino-hydrodynamics code
\textsc{Prometheus-Vertex}. This SN simulation tool combines the
hydrodynamics solver \textsc{Pro\-metheus} with the neutrino transport module
\textsc{Vertex} (see Refs.~\cite{Hanke:2013ena, Tamborra:2014aua} for more
details and additional references). It includes a ``ray-by-ray-plus''
(RbR+), fully velocity and energy-dependent neutrino transport module based
on a variable Eddington-factor technique that solves iteratively the neutrino
energy, momentum, and Boltzmann equations~\cite{Rampp:2002bq, Buras:2005rp}.
We employ state-of-the-art neutrino interaction rates~\cite{Buras:2005rp,
Mueller:2012is} and relativistic gravity and redshift
corrections~\cite{Rampp:2002bq, Marek:2005if}. The RbR+ description assumes
the neutrino momentum distribution to be axisymmetric around the radial
direction everywhere, implying that the neutrino fluxes are radial.

We have performed 3D simulations for the evolution of the 11.2 and
$27\,M_\odot$ progenitors of Woosley, Heger and Weaver \cite{Woosley:2002}
and the 20\,$M_\odot$ model of Woosley and Heger~\cite{Woosley:2007}, using
the high-density equation of state (EoS) of Lattimer and
Swesty~\cite{Lattimer:1991nc} with a nuclear incompressibility of
$K=220$\,MeV. They were previously employed for 2D
simulations~\cite{Marek:2007gr, Mueller:2012is, Mueller:2012ak,
Bruenn:2012mj}. Seed perturbations for aspherical instabilities were imposed
by hand 10\,ms after core bounce by introducing random perturbations of 0.1\%
in density on the entire computational grid  to seed the growth of
hydrodynamic instabilities. None of these models led to successful explosions
during the simulation period of about 350\,ms for the 11.2 and $20\,M_\odot$
models and 550\,ms for the $27\,M_\odot$ case, although explosions were
obtained in the corresponding 2D simulations with the same microphysics. The
postbounce hydrodynamics of the $27\,M_\odot$ model, in particular the
prominent presence of SASI sloshing and spiral modes, was described in a
previous paper~\cite{Hanke:2013ena}, while more details on the hydrodynamics
of the $11.2$ and $20\ M_\odot$ SN progenitors have been provided in our LESA
paper \cite{Tamborra:2014aua}.

The 20 and 27\,$M_\odot$ models both show periods of strong SASI
activity. In the former case, which was simulated until 550\,ms post
bounce (p.b.), a second SASI episode occurs after a period clearly
dominated by
convective overturn. On the other hand, the 11.2\,$M_\odot$ model does
not exhibit any clear evidence of SASI motions but develops the
typical signatures of postshock convective overturn in the
neutrino-heating layer.

We will usually show neutrino flux characteristics as they would be
seen by a distant observer located at chosen angular coordinates in
the coordinate system of the SN simulation.  For any angular position,
the neutrino luminosity reaching the observer is given by the
superposition of the projected fluxes emitted under different
angles, as described in Appendix~\ref{sec:fluxprojections}. Therefore,
the observable neutrino fluxes are weighted hemispheric averages performed such as to
include flux projection effects in the observer direction. The
hemispheric averages, as expected, show smaller time variations than
specific angular rays.

\begin{figure}
\centering
\includegraphics[width=0.9\columnwidth]{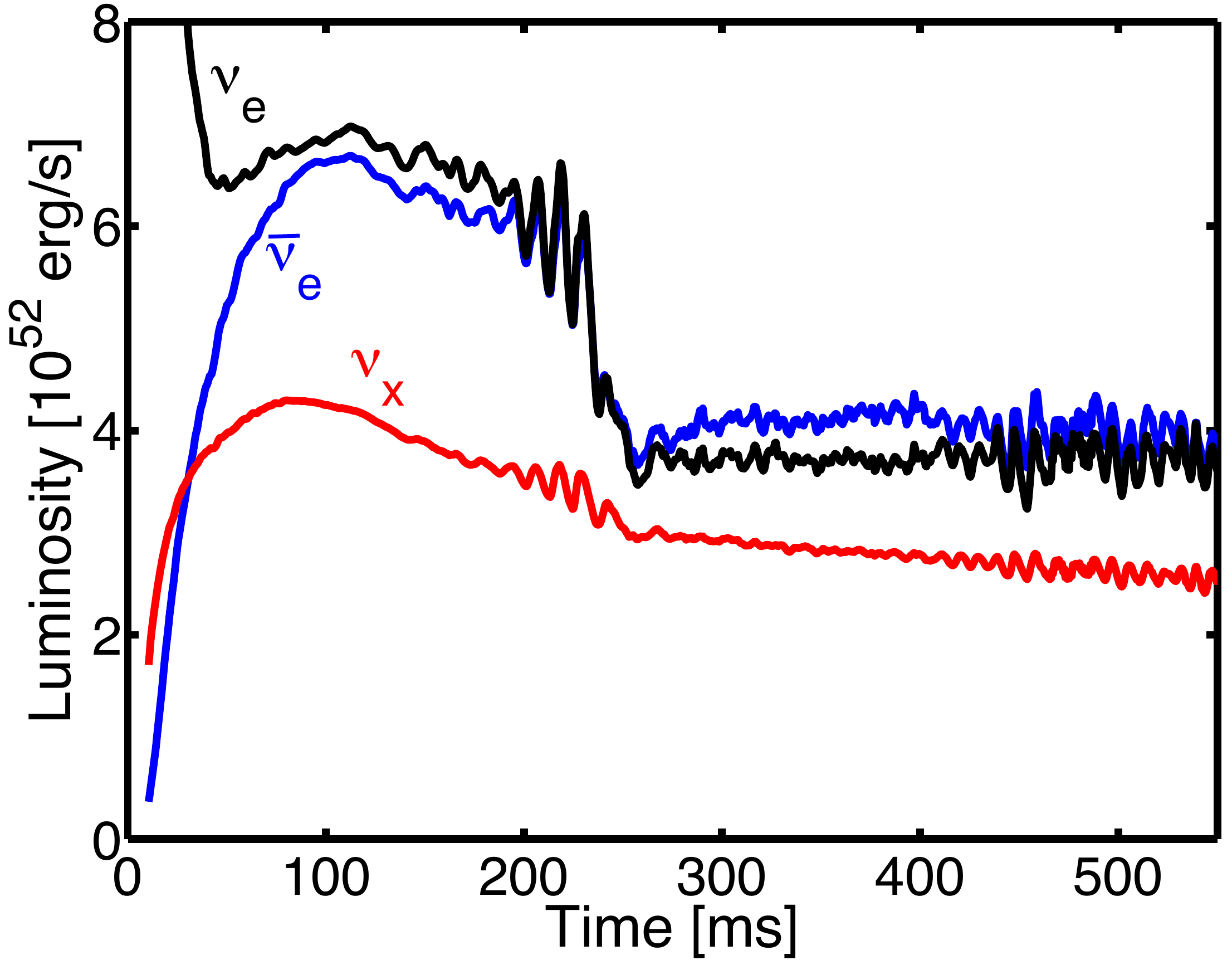}
\caption{Luminosity of the $\nu_e$, $\bar{\nu}_e$ and $\nu_x$ species
  for our $27\,M_\odot$ simulation as measured by a distant observer
  with angular coordinates close to the plane of the spiral mode in the
  first SASI period.
\label{fig:luminosities}}
\end{figure}

As a benchmark example, we show in Fig.~\ref{fig:luminosities} the luminosity
for $\nu_e$, $\bar\nu_e$ and $\nu_x=\nu_{\mu}$, $\nu_{\tau}$, $\bar\nu_{\mu}$
or $\bar\nu_{\tau}$ as a function of time, as seen by a distant observer with
angular coordinates close to the plane of the SASI spiral mode.
Large-amplitude, near-sinusoidal modulations of the neutrino signal occur in
the interval 120--260~ms as imprinted by SASI. For 260--410~ms  
a convective phase occurs, followed by another SASI episode on a
different plane with respect to the previous one.  SASI modulations have a
similar amplitude for $\nu_e$ and $\bar{\nu}_e$, while they are somewhat
smaller for $\nu_x$.

\begin{figure}
\centering
\includegraphics[width=0.709\columnwidth]{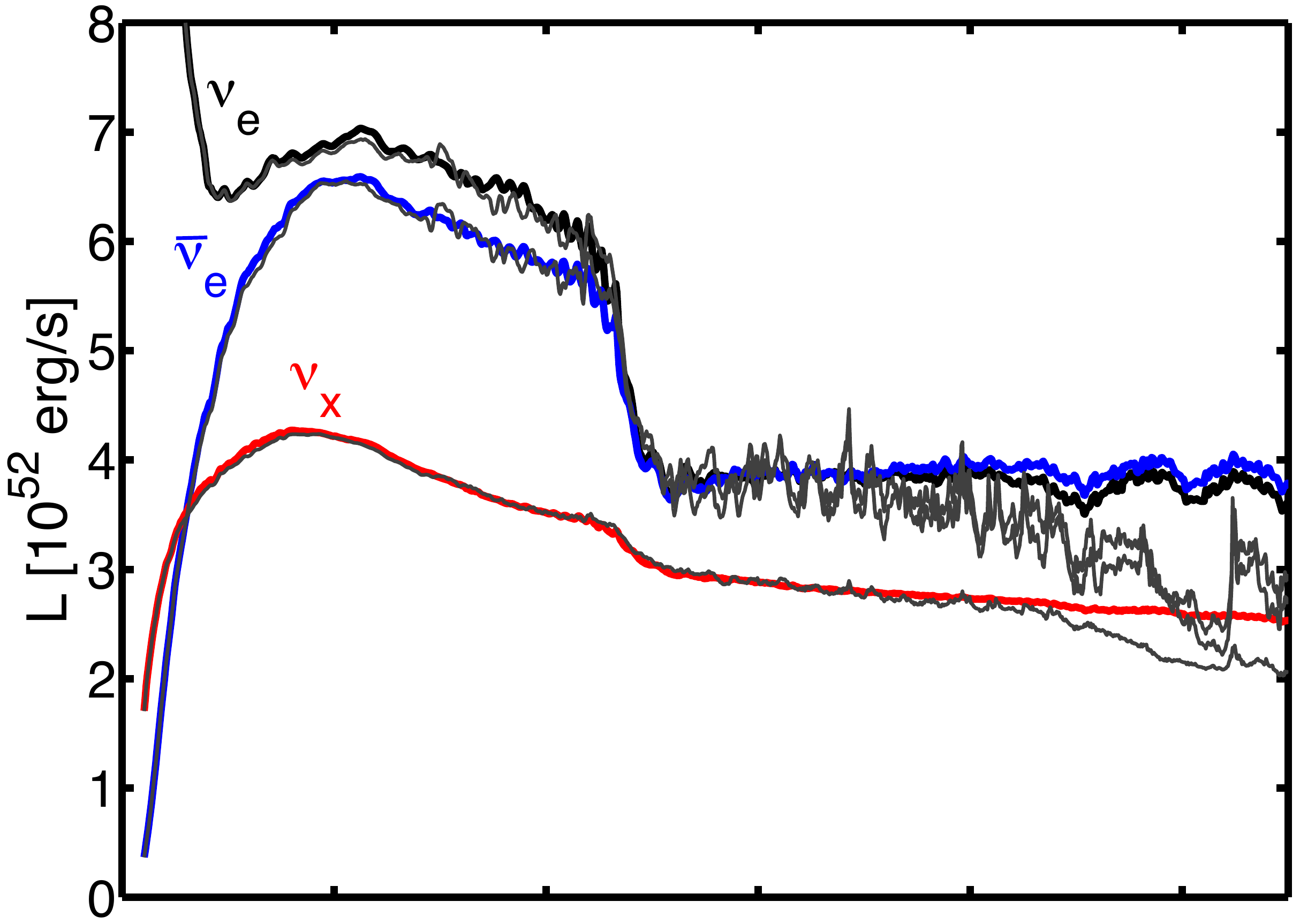}
\hspace*{-1.mm}\includegraphics[width=0.718\columnwidth]{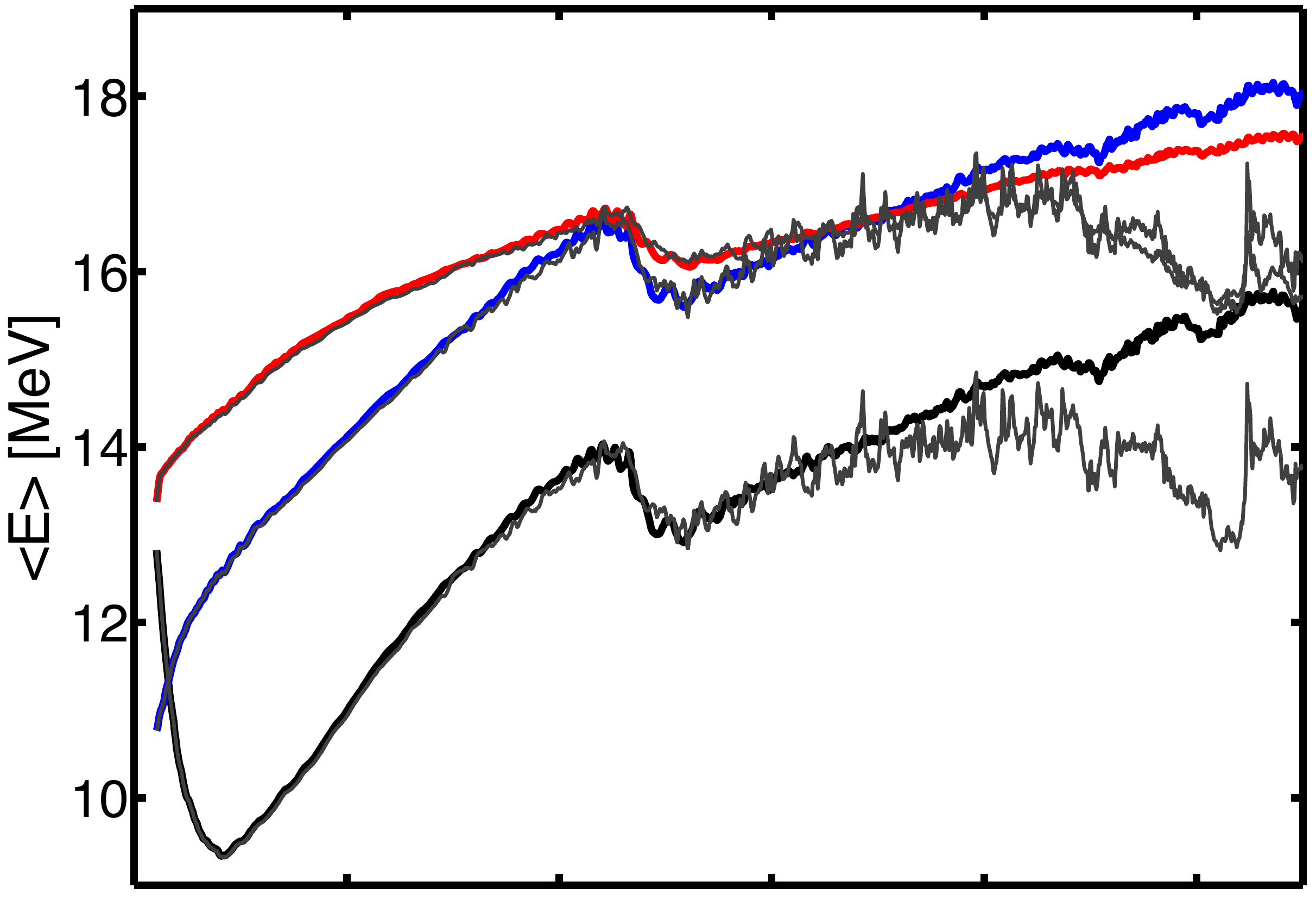}
\hspace*{-1mm}\includegraphics[width=0.718\columnwidth]{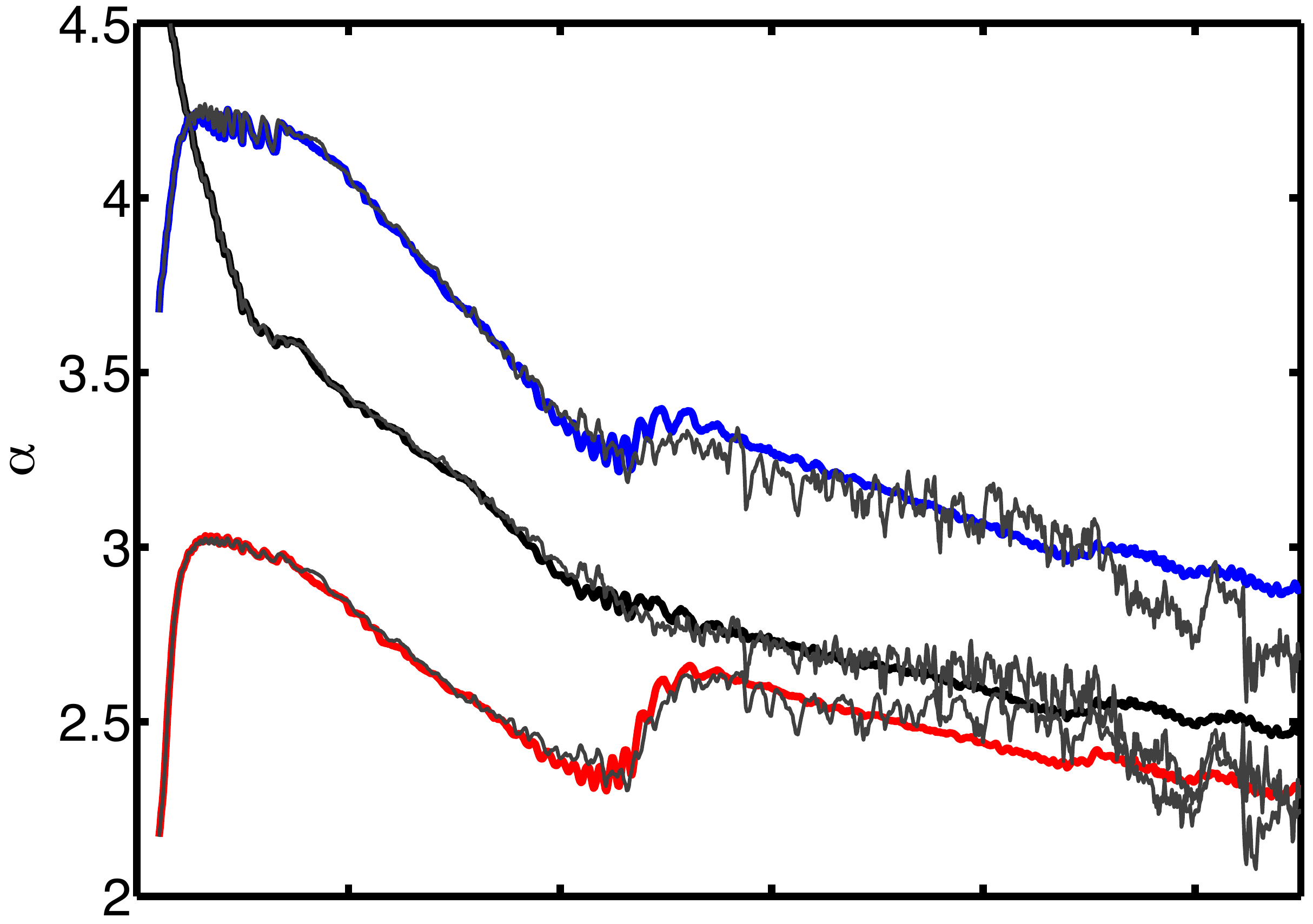}
\hspace*{-2mm}\includegraphics[width=0.738\columnwidth]{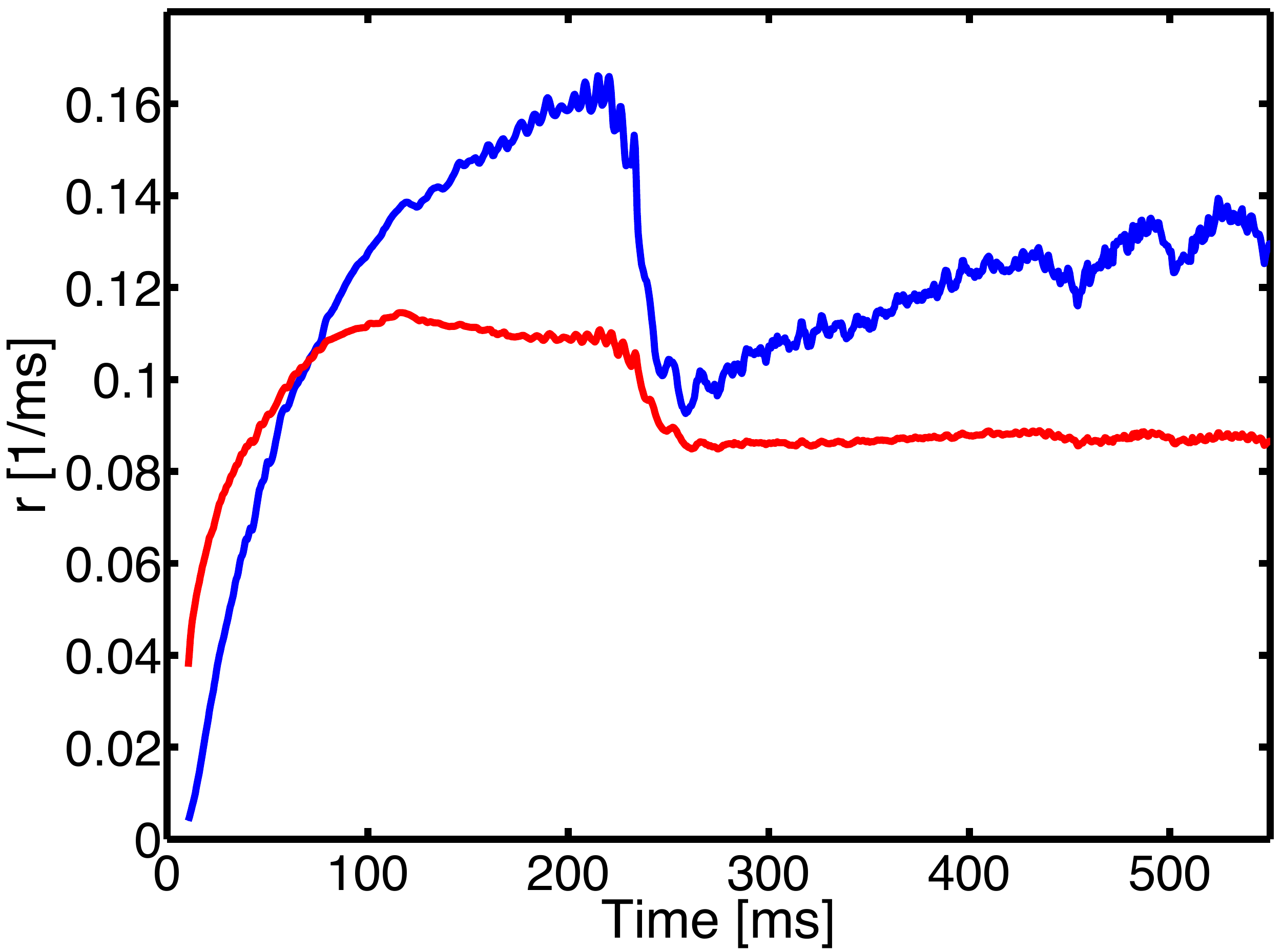}
\caption{Neutrino flux properties of our $27\,M_\odot$ case after
  integrating over all directions.  For $\nu_e$, $\bar\nu_e$ and
  $\bar\nu_x$ we show the luminosity, average energy and shape
  parameter $\alpha$ from 3D (in black, blue and red respectively) and
  2D (in grey) simulations for comparison.  The single-OM IceCube rate $r$ in the
  bottom panel is without dead time for a SN distance of 10~kpc. Blue
  line: based on $\bar\nu_e$ flux without flavor oscillations. Red
  line: based on $\bar\nu_x$, i.e., assuming full flavor swap
  $\bar\nu_e\leftrightarrow\bar\nu_e$.\label{fig:averageSN}}
\end{figure}

Figure~\ref{fig:averageSN} shows the properties of our $27\,M_\odot$
simulation, averaged over all directions, to mimic an equivalent spherically
symmetric case. Of course, this average does not depend on observer-related
projection effects.  For the species $\nu_e$, $\bar\nu_e$ and $\nu_x$, we
show the luminosity, average energy, and shape parameter $\alpha$ of the
assumed spectral 
Gamma distribution (Appendix~\ref{sec:ibd}). The fast time
variations here have very small amplitude, i.e., convection and SASI activity
do not strongly modulate the overall neutrino emission parameters---the
modulations in various directions essentially cancel out.

The hierarchy of fluxes and average energies as well as the shape parameter
correspond to what is expected. It is noteworthy, however, that the average
$\bar\nu_e$ and $\bar\nu_x$ energies become very similar after around 220~ms,
at the end of the first SASI episode, when the shock wave has considerably
expanded. This feature has been seen in previous simulations~\cite{Marek:2008qi,Mueller:2014rna}
too, and reflects the temperature increase in the settling, growing accretion layer on the
proto-neutron star core. This accretion layer radiates mainly $\nu_e$
and $\bar\nu_e$ and downgrades the $\nu_x$ escaping from deeper layers
in energy space~\cite{Keil:2002in}.
The pronounced luminosity drop at $\sim$250\,ms occurs
because of the infall of the Si/SiO shell interface leading to strong shock
expansion and therefore to a dramatic decrease of the mass accretion
rate~\cite{Hanke:2013ena}.

Although the difference between 2D and 3D models is not subject of our work,
in Fig.~\ref{fig:averageSN} we show the corresponding 2D spectral parameters
averaged over all directions. The 3D and 2D integrated quantities are very
similar up to 300~ms when the 2D model explodes.

\begin{figure}
\centering
\includegraphics[width=0.9\columnwidth]{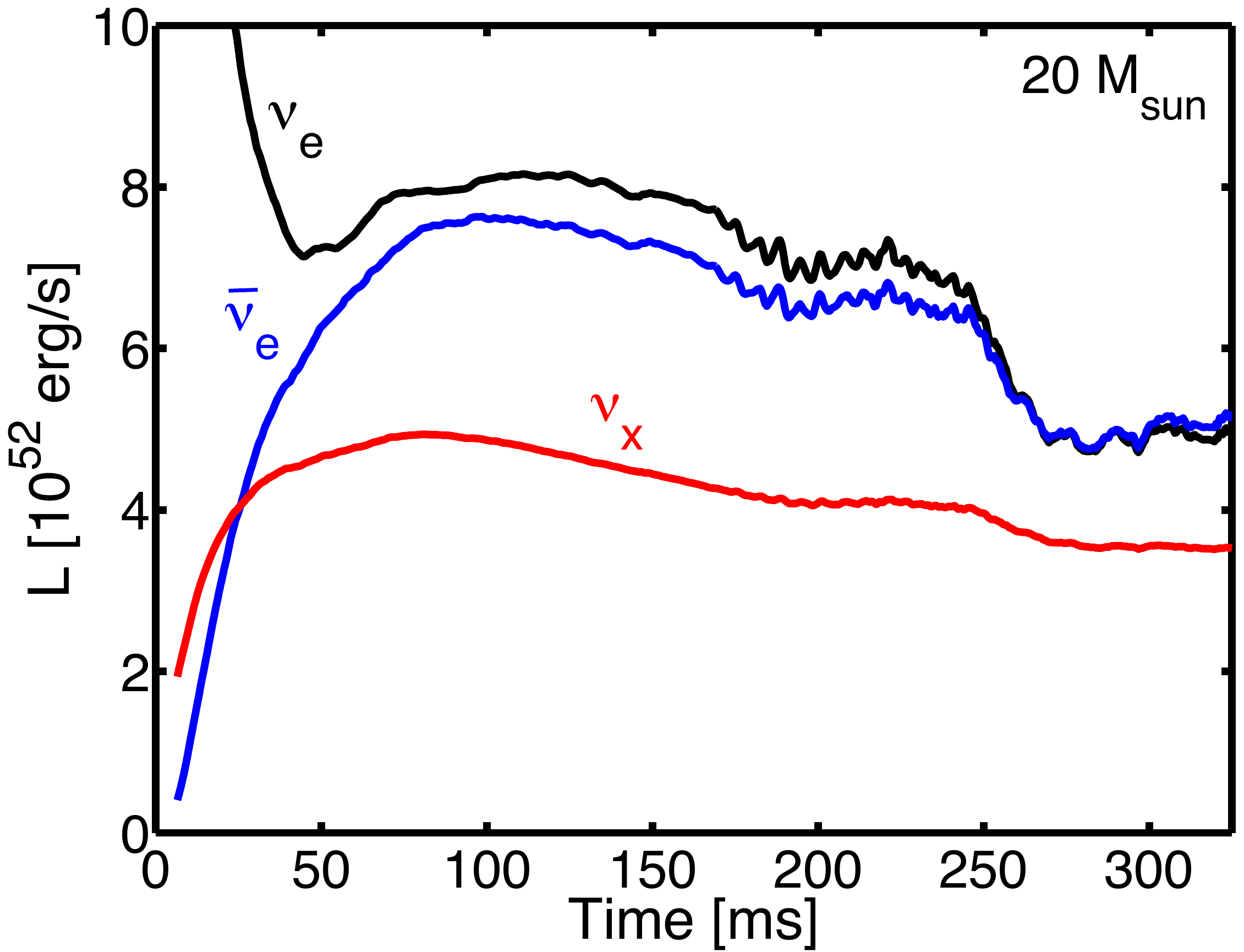}
\caption{Luminosity of the $\nu_e$, $\bar{\nu}_e$ and $\nu_x$ species
  for our $20\,M_\odot$ simulation after
  integrating over all directions.
\label{fig:luminosities20}}
\end{figure}

Figure~\ref{fig:luminosities20} shows the luminosity of
the $\nu_e$, $\bar{\nu}_e$ and $\nu_x$ species for our $20\,M_\odot$
simulation, averaged over all directions to mimic an equivalent
spherically symmetric case, for comparison with the top panel of
Fig.~\ref{fig:averageSN}. The hierarchy among the luminosities of
different flavors as well as their behavior as a function of time is
similar for both the 20 and $27\,M_\odot$ progenitors.  However, the
luminosities of $\nu_e$ and $\bar{\nu}_e$ are slightly higher for the
$20\,M_\odot$ simulation. Despite the average over all directions, the
integrated luminosities present residual sinusoidal modulations for
$t \ge 160$~ms (i.e., during the SASI episode) with amplitude larger
than for the $27\,M_\odot$ simulation (see Fig.~\ref{fig:averageSN},
top panel) because the SASI activity is stronger for this SN model.

\section{Neutrino Signal from 3D Models}
\label{sec:analysis}

\subsection{\boldmath{$11\,M_\odot$} progenitor}
\label{sec:11Msun}

\begin{figure}
\centering
\includegraphics[width=0.9\columnwidth]{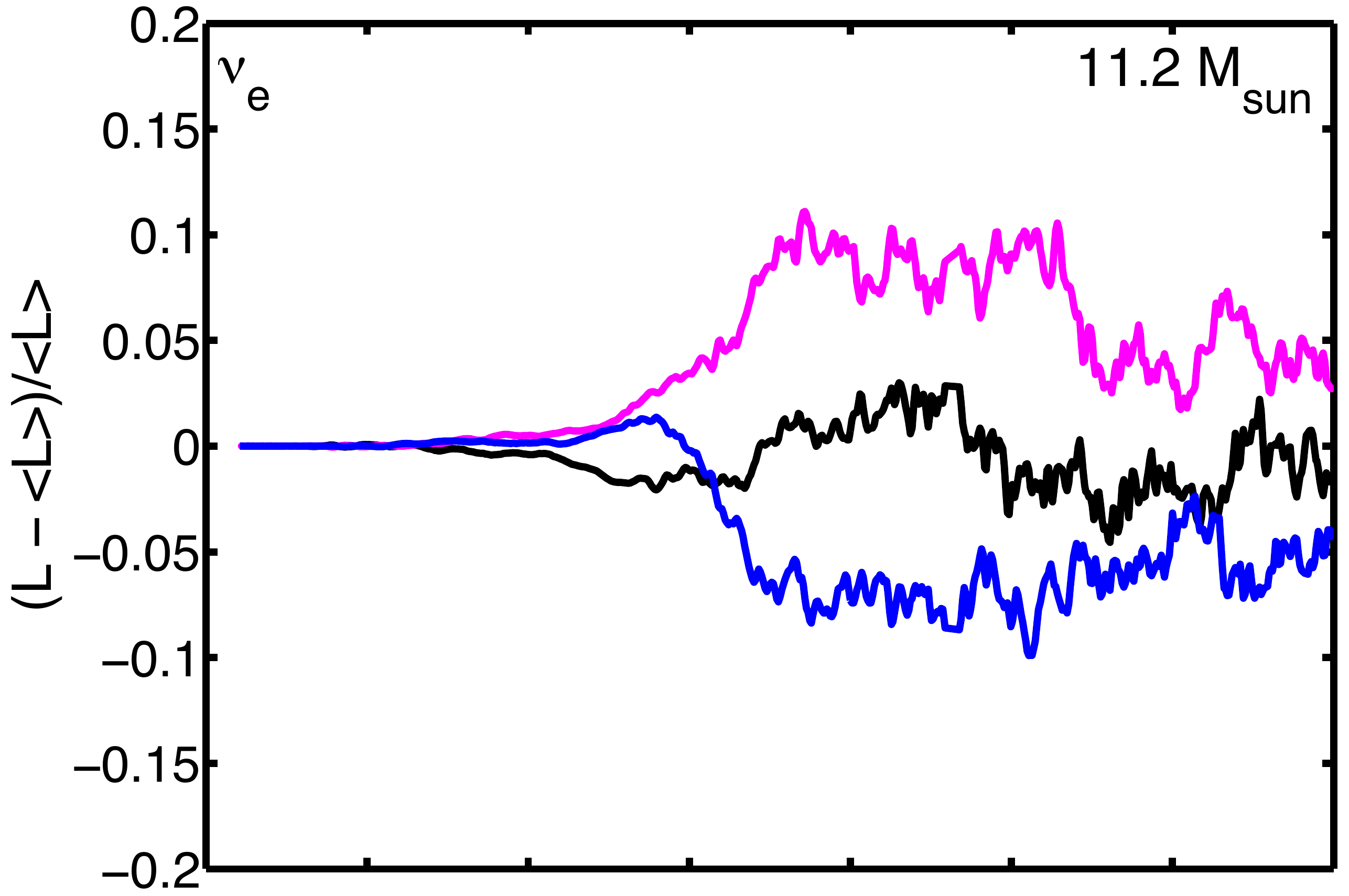}\\
\includegraphics[width=0.9\columnwidth]{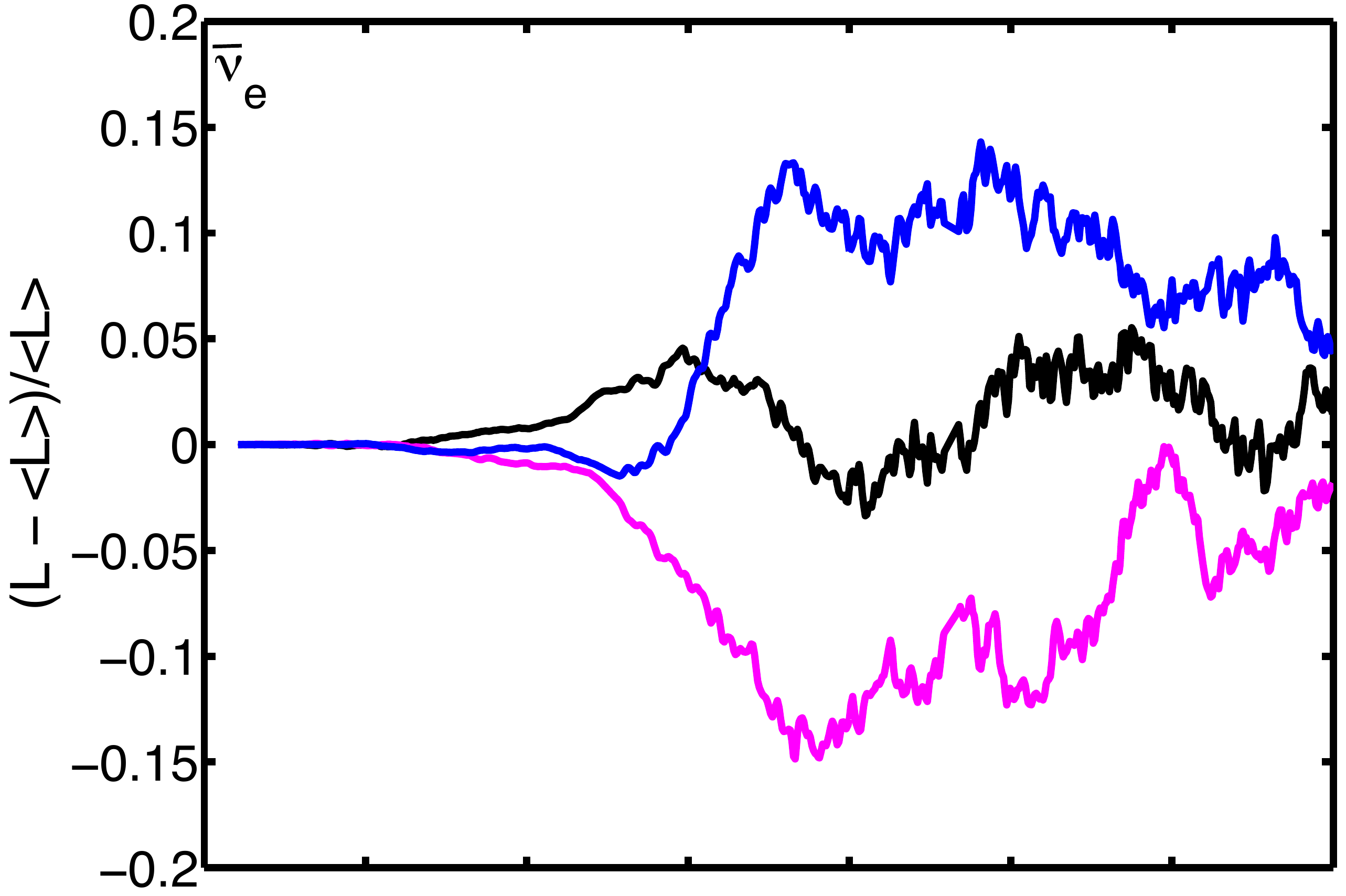}\\
\hspace{2mm}\includegraphics[width=0.92\columnwidth]{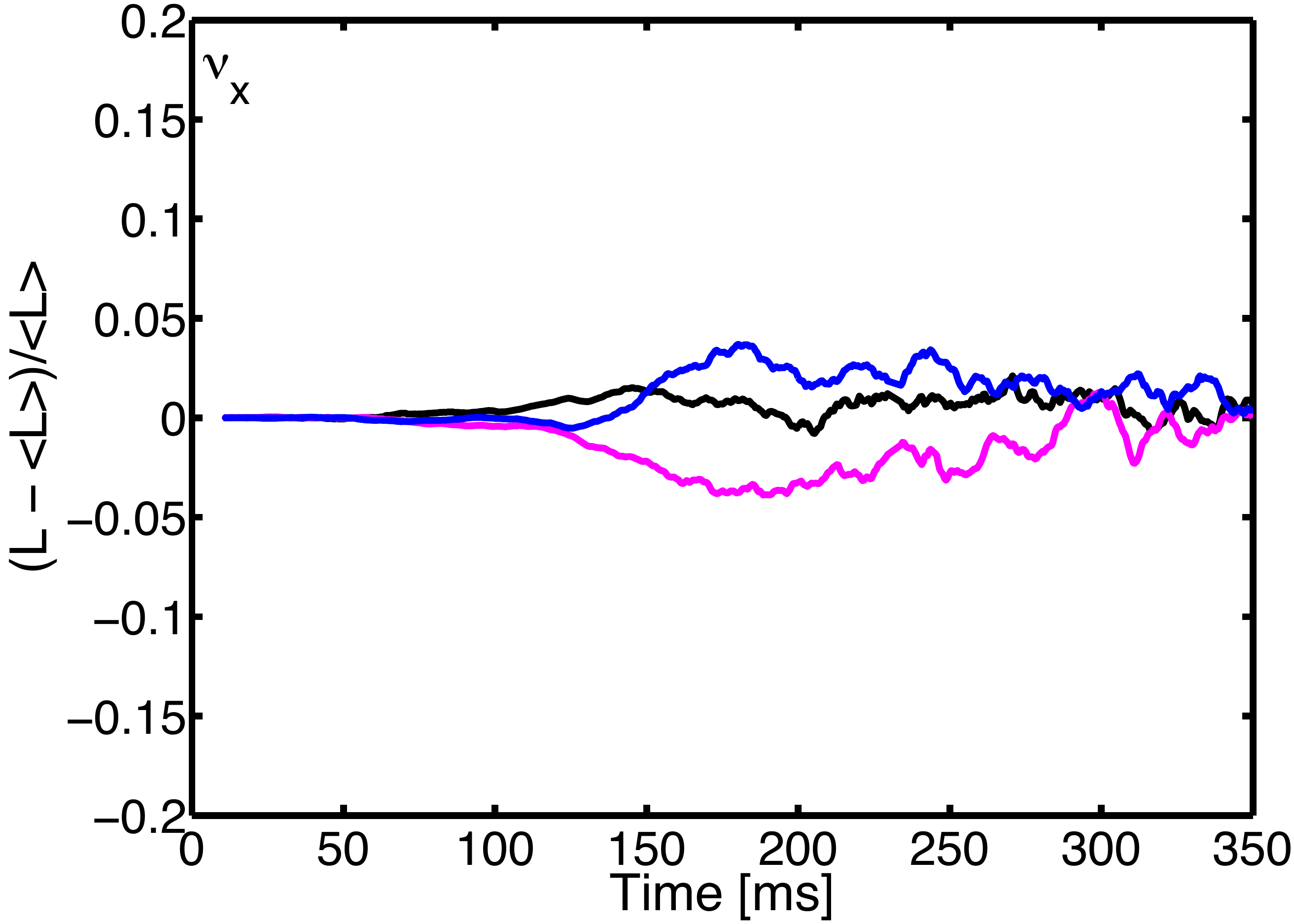}
\caption{Luminosity evolution for the $11.2\ M_\odot$ progenitor, separately
  for $\nu_e$, $\bar{\nu}_e$ and $\nu_x$, as
  seen by a distant observer, relative to $\langle L \rangle$, the
  time-dependent average over all directions.
  Blue and magenta curves: Observer location on opposite sides along
  the axis where the flux variations are largest. Black curve: One
  typical orthogonal direction where the variation is small. The large
  excursion of the blue and magenta lines represent the LESA
  phenomenon: The $\nu_e$ and $\bar\nu_e$ emission show a strong
  dipole pattern. The small-amplitude fast time variations are caused
  by large-scale convective overturn. There is no SASI activity in
  this model.\label{fig:lum11}}
\end{figure}

We now turn to a detailed discussion of the direction and time
dependent features of the observable neutrino signal emitted by our 3D
models. Beginning with the $11\,M_\odot$ progenitor,
Fig.~\ref{fig:lum11} shows the luminosity evolution, $L$, relative to
the time-dependent average $\langle L \rangle$ over all directions,
separately for $\nu_e$, $\bar{\nu}_e$ and $\nu_x$.  This model does
not show any SASI activity, but only small-amplitude, fast time
variation caused by large-scale convective overturn. However after
some 150~ms, the $\nu_e$ and $\bar\nu_e$ luminosities develop a
quasi-stationary dipole pattern, representing the LESA effect
discussed in our earlier paper~\cite{Tamborra:2014aua}.

\begin{figure*}[p]
\centering
\includegraphics[width=0.95\textwidth]{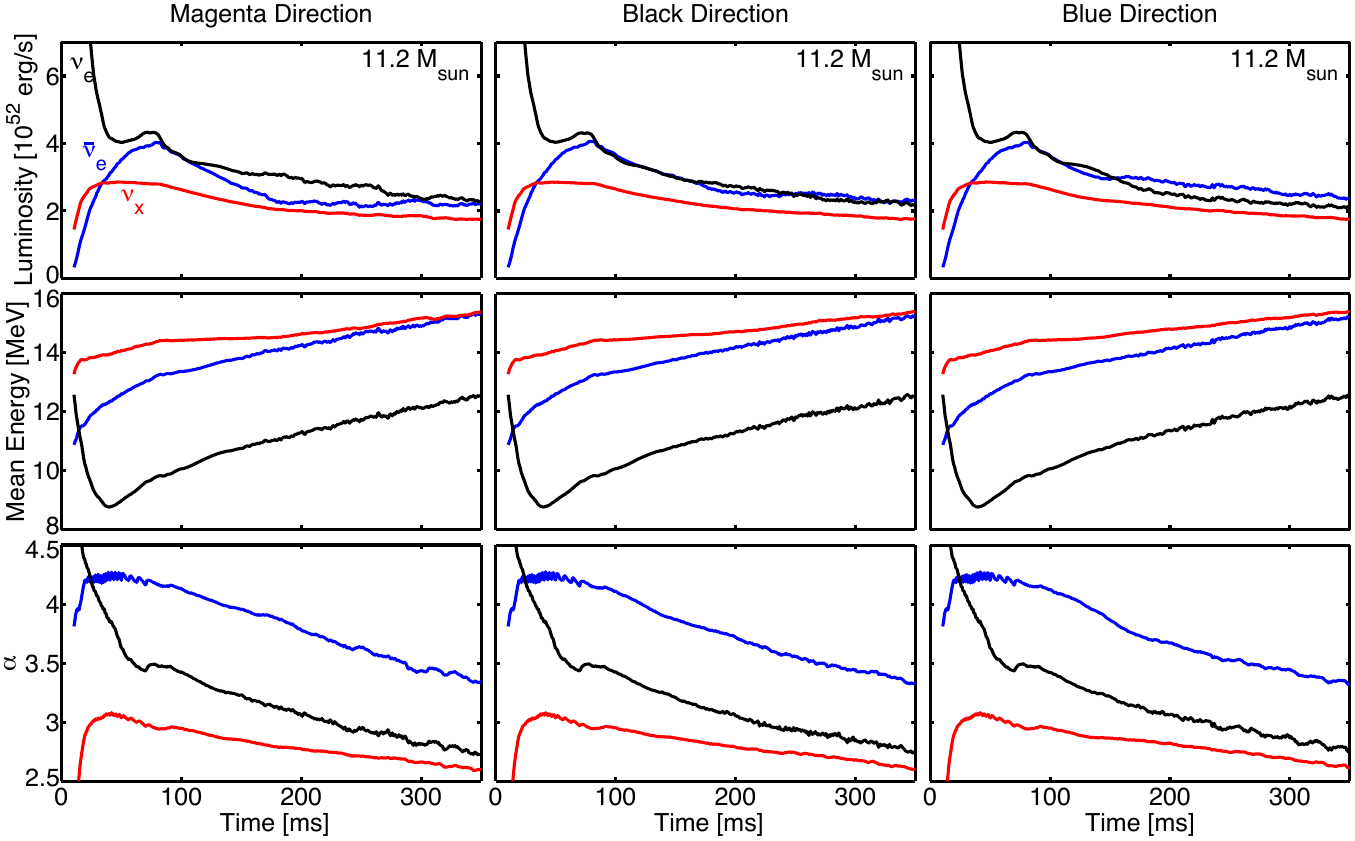}
\caption{Evolution of neutrino flux properties for the
  $11.2\ M_{\odot}$ progenitor as seen from a distant observer. For
  $\nu_e$, $\bar{\nu}_e$ and $\nu_x$ we show the luminosity, average
  energy and shape parameter $\alpha$. The ``Magenta'' and ``Blue'' directions
  are opposite
  along the LESA axis, corresponding to the magenta and blue curves in
  Fig.~\ref{fig:lum11}, whereas the
  ``Black'' direction is on the LESA equator (black in Fig.~\ref{fig:lum11}).
\label{fig:goodbad_dir_11}}
\vskip20pt
\centering
\includegraphics[width=0.95\textwidth]{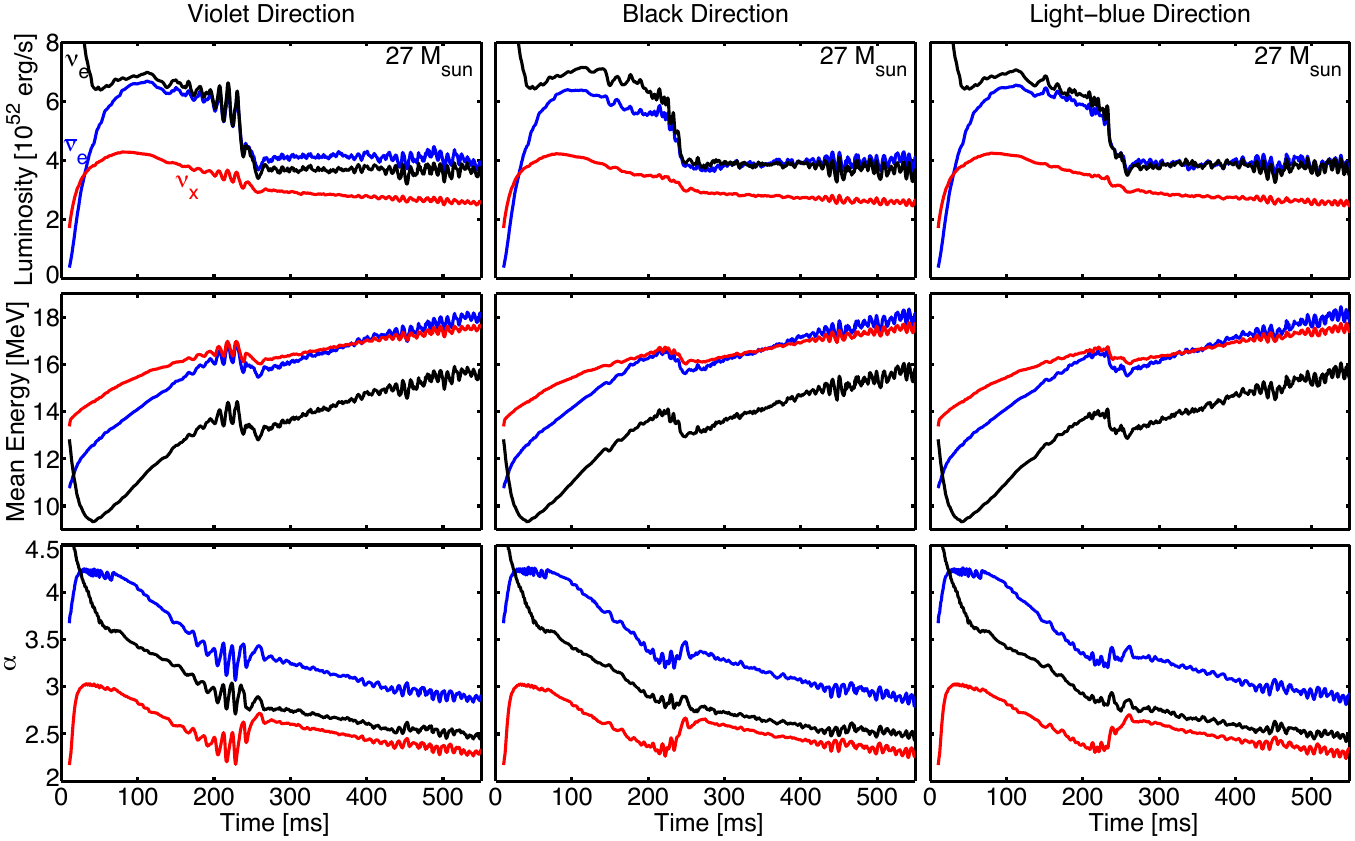}
\caption{Same as Fig.~\ref{fig:goodbad_dir_11}, but
for the $27\,M_{\odot}$ progenitor. The ``Violet,'' ``Black''
and ``Light Blue'' directions here correspond to the curves of the same color
in Fig.~\ref{fig:lum27} that were chosen to show large and small SASI
amplitudes, respectively. \label{fig:goodbad_dir}}
\end{figure*}

The two observer directions shown in Fig.~\ref{fig:lum11} (blue and magenta
lines) are chosen on opposite sides of the SN along the LESA axis. The black
curve represents a typical orthogonal direction, i.e., it is on the ``LESA
equator.'' The observer directions remain fixed in time whereas the LESA
dipole direction slowly drifts, so in this sense these directions are not
always exactly along the LESA axis or equator, respectively.

As discussed in our LESA paper~\cite{Tamborra:2014aua}, the sum of
all flavor luminosities is almost independent of direction
and the
$\bar\nu_e$ and $\nu_e$ dipoles point in opposite directions. However, in a
realistic detector, we measure only the $\bar\nu_e$ signal by inverse beta
decay. Ignoring flavor oscillations,  the measurable $L_{\bar\nu_e}$ could
therefore differ by as much as 30\% during the accretion phase, depending on
the observer location, affecting the implied overall neutrino luminosity. Of
course, what is really measured in a detector depends on flavor conversion
which likely is a large effect. Since the $\bar\nu_x$ fluxes show a much
weaker directional modulation, the real uncertainty between the measurement
and the true $4\pi$ equivalent flux will be less dramatic.

In order to quantify the directional dependence of the neutrino signal,
Fig.~\ref{fig:goodbad_dir_11} shows the neutrino flux properties (luminosity,
mean energy and the shape parameter $\alpha$) for the three species along the
same three directions chosen in Fig.~\ref{fig:lum11}, i.e., ``Magenta,''
``Black'' and ``Blue'' directions respectively named by the curves of the
same colors shown in Fig.~\ref{fig:lum11}. We recall that the flux
characteristics pertain to observers in those directions, i.e., they involve
hemispheric averaging with appropriate flux projections. The small-amplitude
``vibrations'' of these parameters are caused by accretion variations
associated with convective overturn.

The hierarchy among flavor-dependent luminosities along the three
directions is slightly different. In particular,
$L_{\nu_e}>L_{\bar{\nu}_e}$ along the ``Magenta'' direction, they are almost
comparable along the ``Black'' direction, and $L_{\nu_e} < L_{\bar{\nu}_e}$ along
the ``Blue'' direction, while the remaining neutrino flux properties exhibit the
same hierarchy independently of the observer direction.

\subsection{20 and \boldmath{$27\,M_\odot$} progenitors}
\label{sec:2027Msun}

\begin{figure}
\centering
\hspace{-1.5mm}\includegraphics[width=0.915\columnwidth]{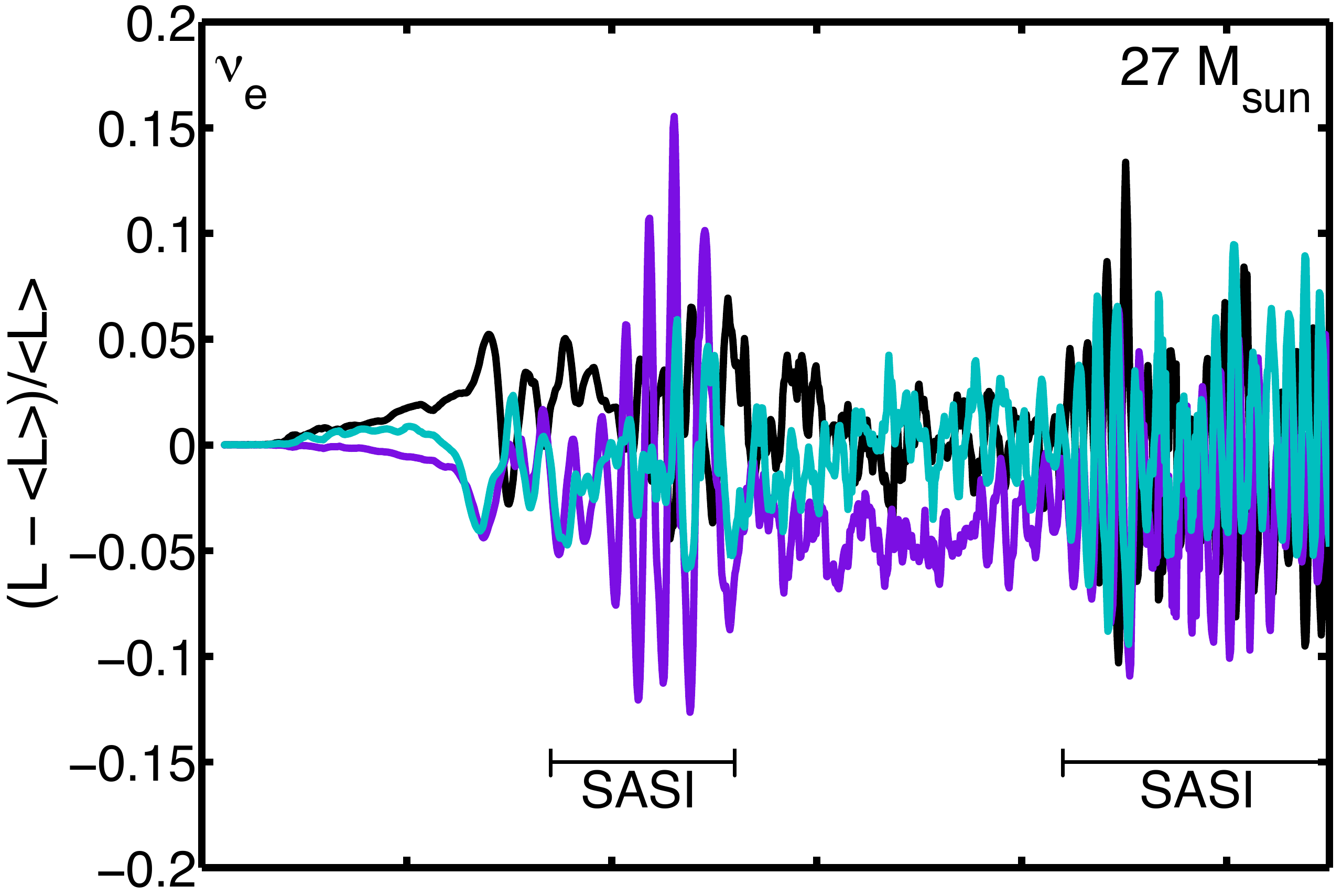}\\
\hspace{-1.5mm}\includegraphics[width=0.92\columnwidth]{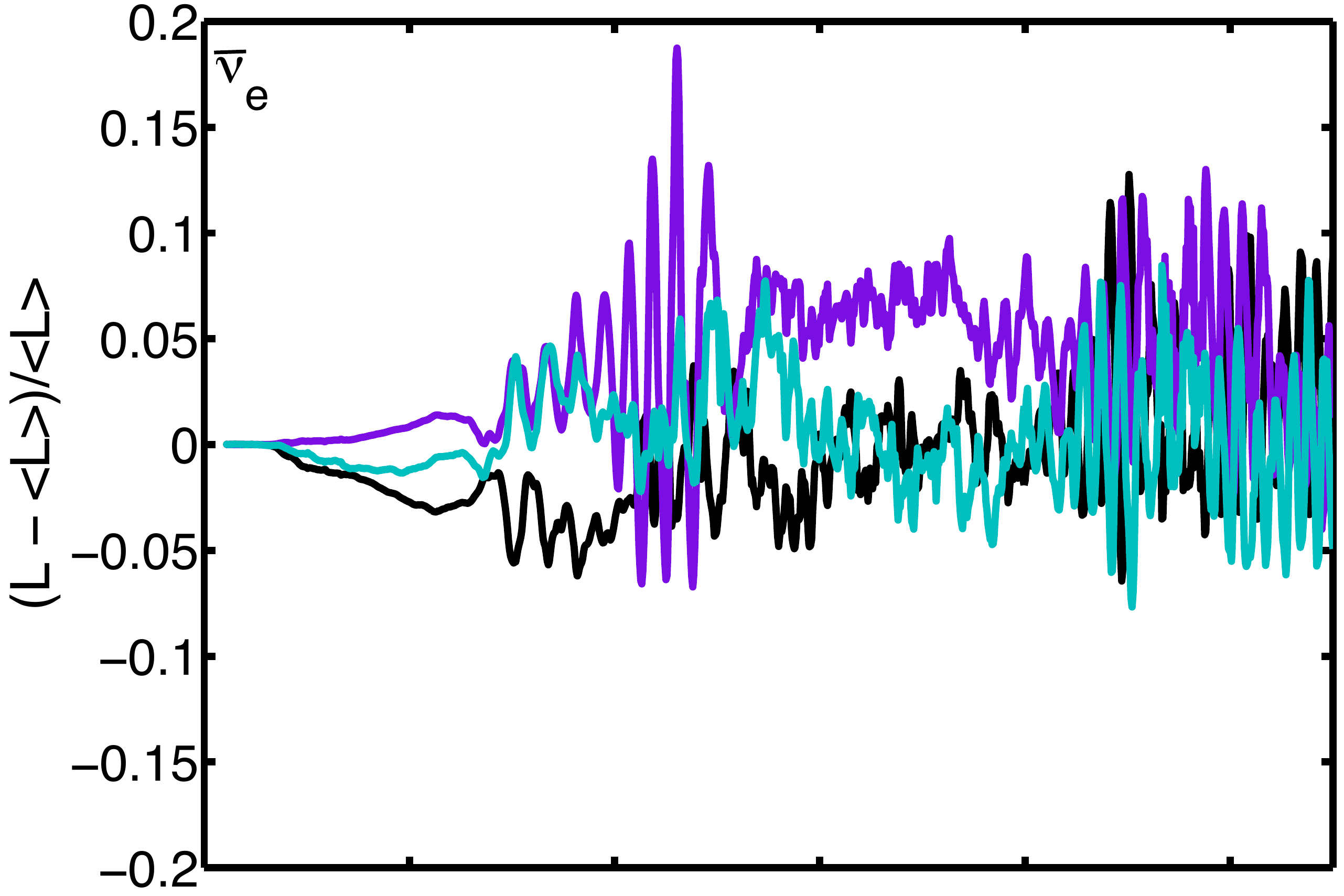}\\
\includegraphics[width=0.915\columnwidth]{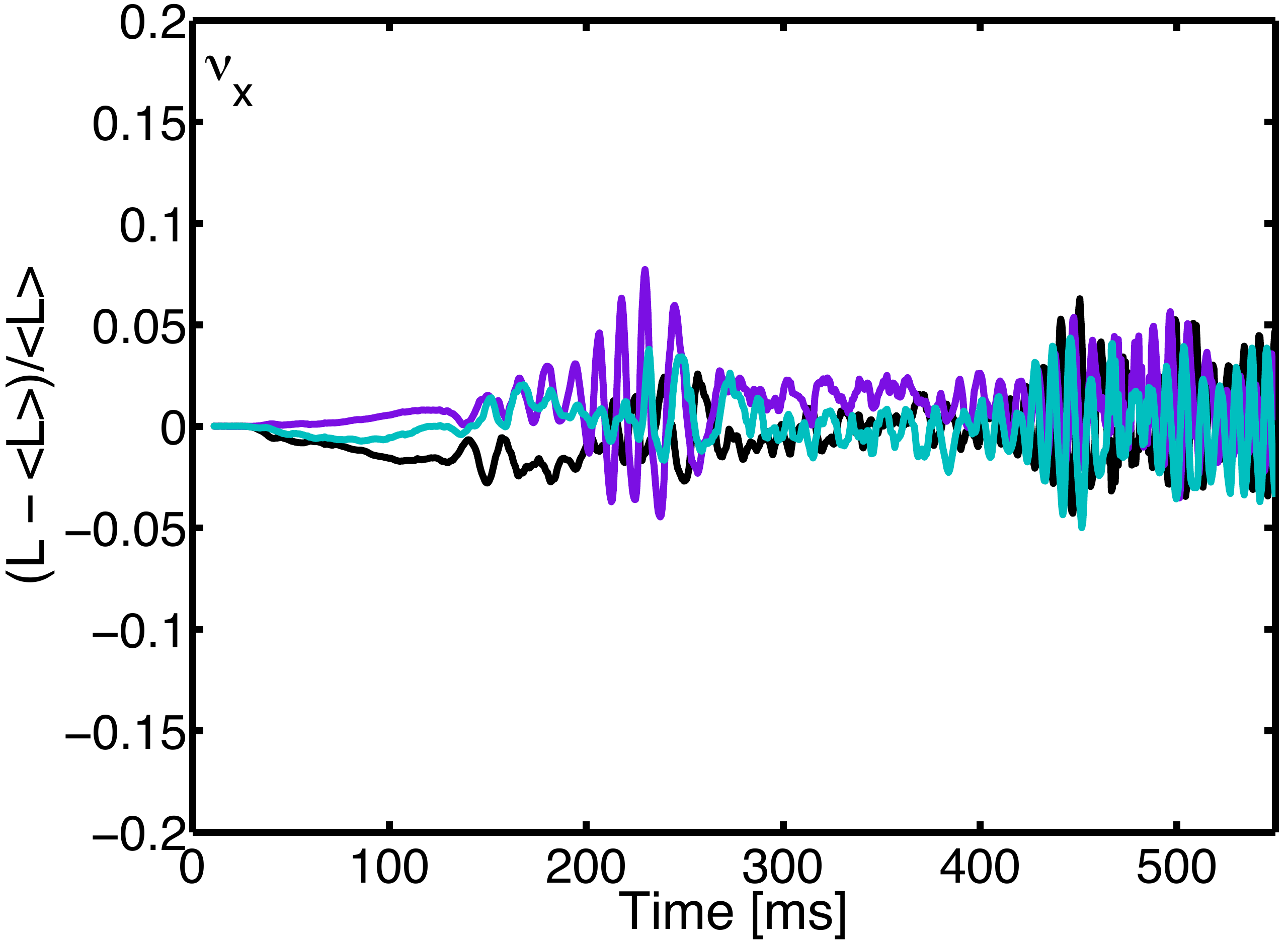}
\caption{Luminosity evolution for the $27\,M_\odot$ progenitor, separately for
  $\nu_e$, $\bar{\nu}_e$ and $\nu_x$, as
  seen by a distant observer, relative to $\langle L \rangle$, in analogy
  to Fig.~\ref{fig:lum11}. The light blue and violet curves refer to observer
  locations in opposite directions approximately within the plane where SASI
  develops. The black line refers to a location of the observer far from the SASI plane
  where
the modulation of the neutrino signal due to SASI is smaller during the first
SASI episode.
\label{fig:lum27}}
\end{figure}

\begin{figure}
\centering
\hspace{-1.5mm}\includegraphics[width=0.915\columnwidth]{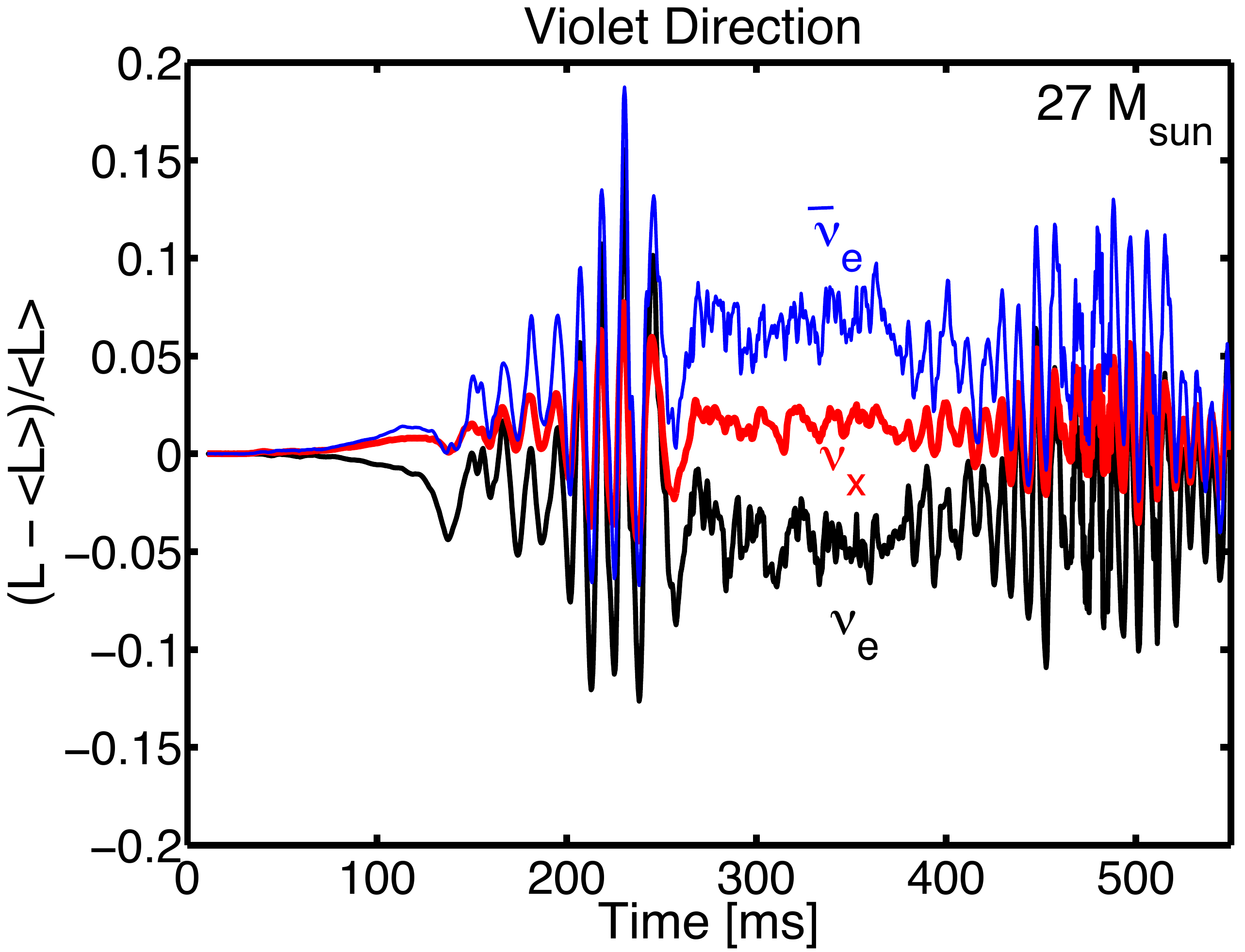}
\caption{Luminosity evolution relative to $\langle L \rangle$ for the $27\,M_\odot$
progenitor for the three species, as seen by a distant observer,
along the ``Violet'' direction, corresponding
to the curve of this color in  Fig.~\ref{fig:lum27}.
Between SASI episodes, we see clear evidence for the LESA asymmetry,
although the chosen direction does not coincide with the LESA axis.
\label{fig:violet27}}
\end{figure}

\begin{figure}
\centering
\includegraphics[width=0.9\columnwidth]{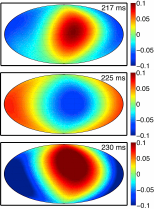}
\caption{Sky maps of $L_{\bar\nu_e}$ relative to the $4\pi$ average
for the $27\ M_\odot$ SN progenitor at $t=217$, 225 and 230~ms,
corresponding to subsequent SASI maxima and minima.
\label{fig:lum27times}}
\end{figure}

\begin{figure}
\centering
\includegraphics[width=0.9\columnwidth]{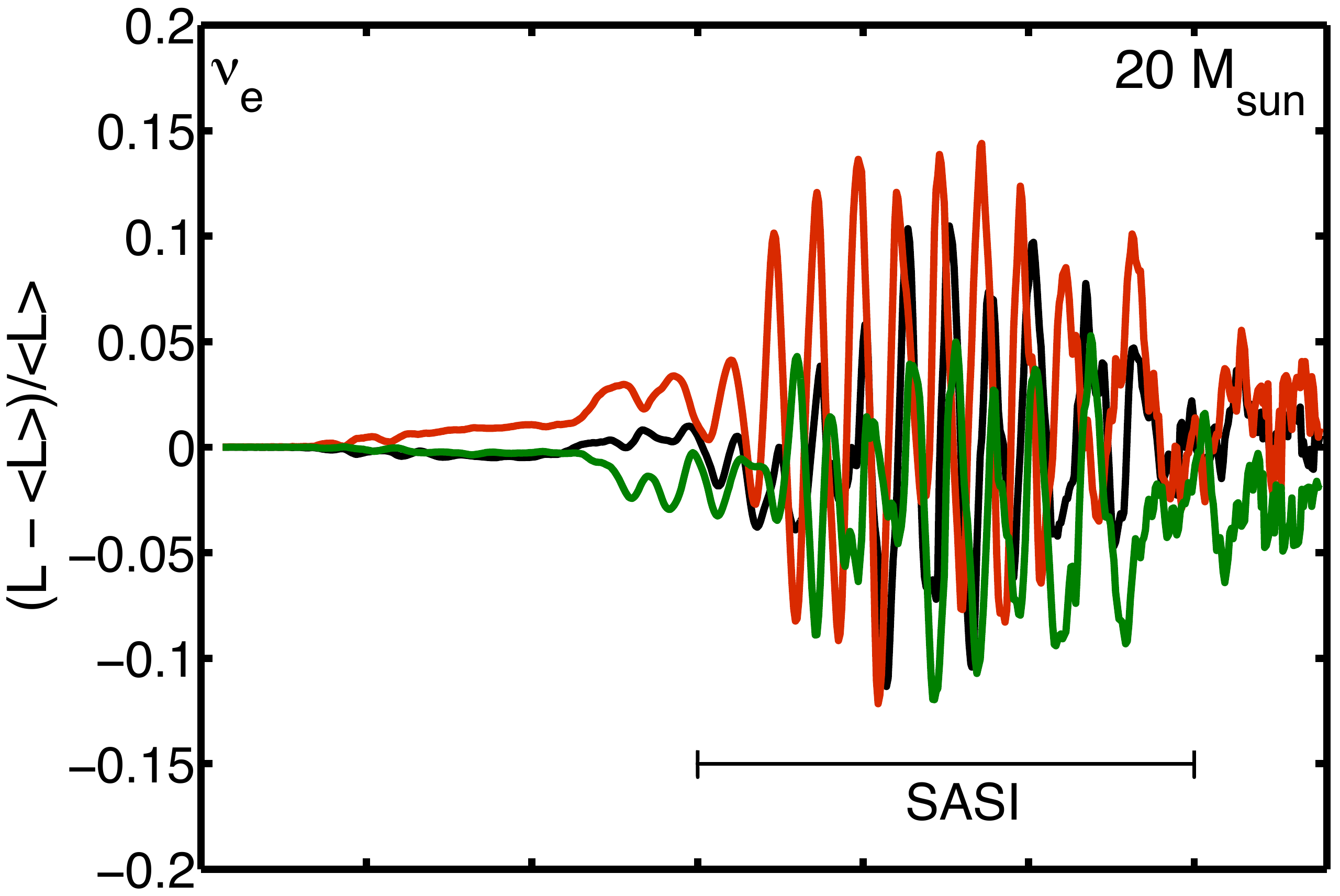}\\
\includegraphics[width=0.9\columnwidth]{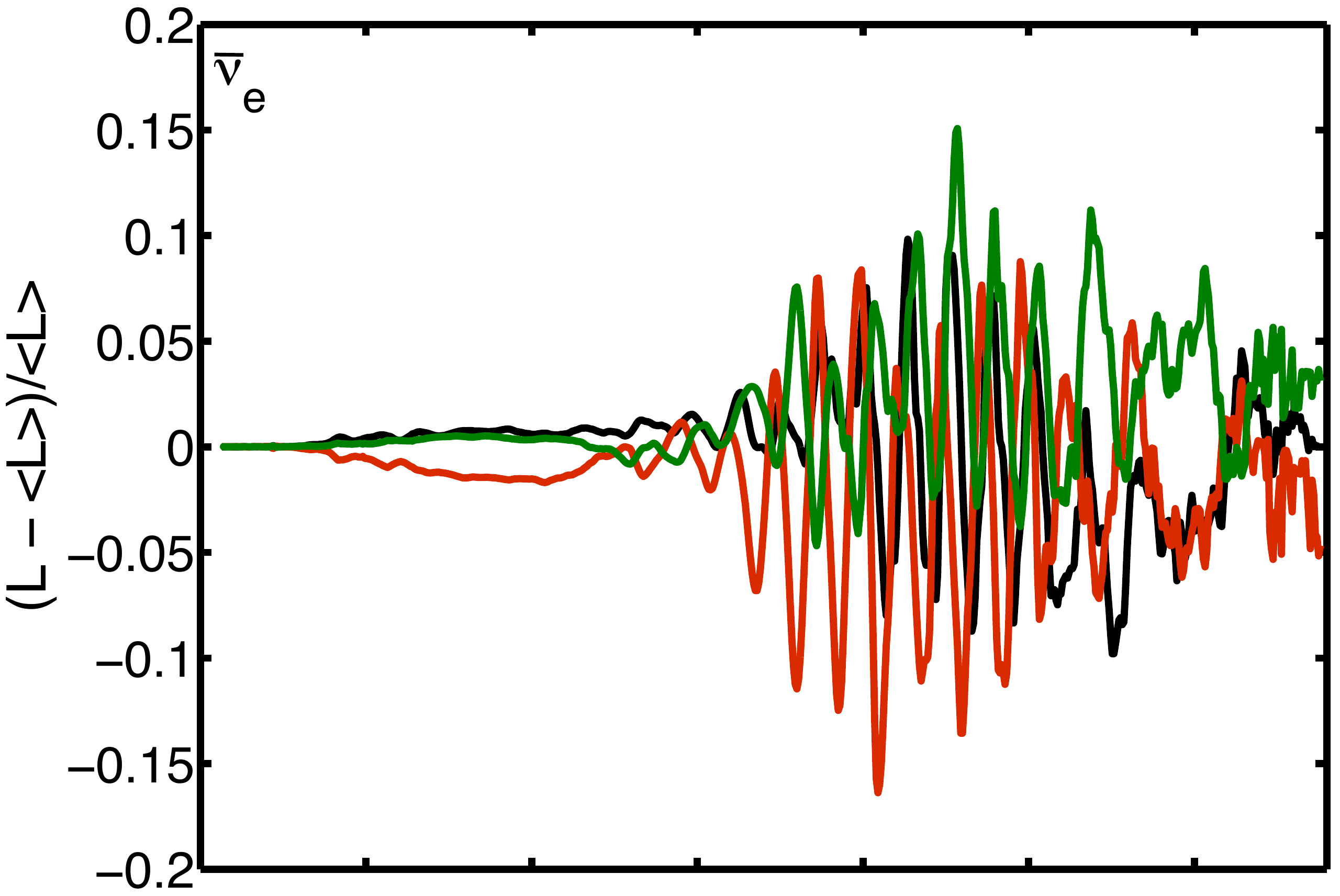}\\
\includegraphics[width=0.9\columnwidth]{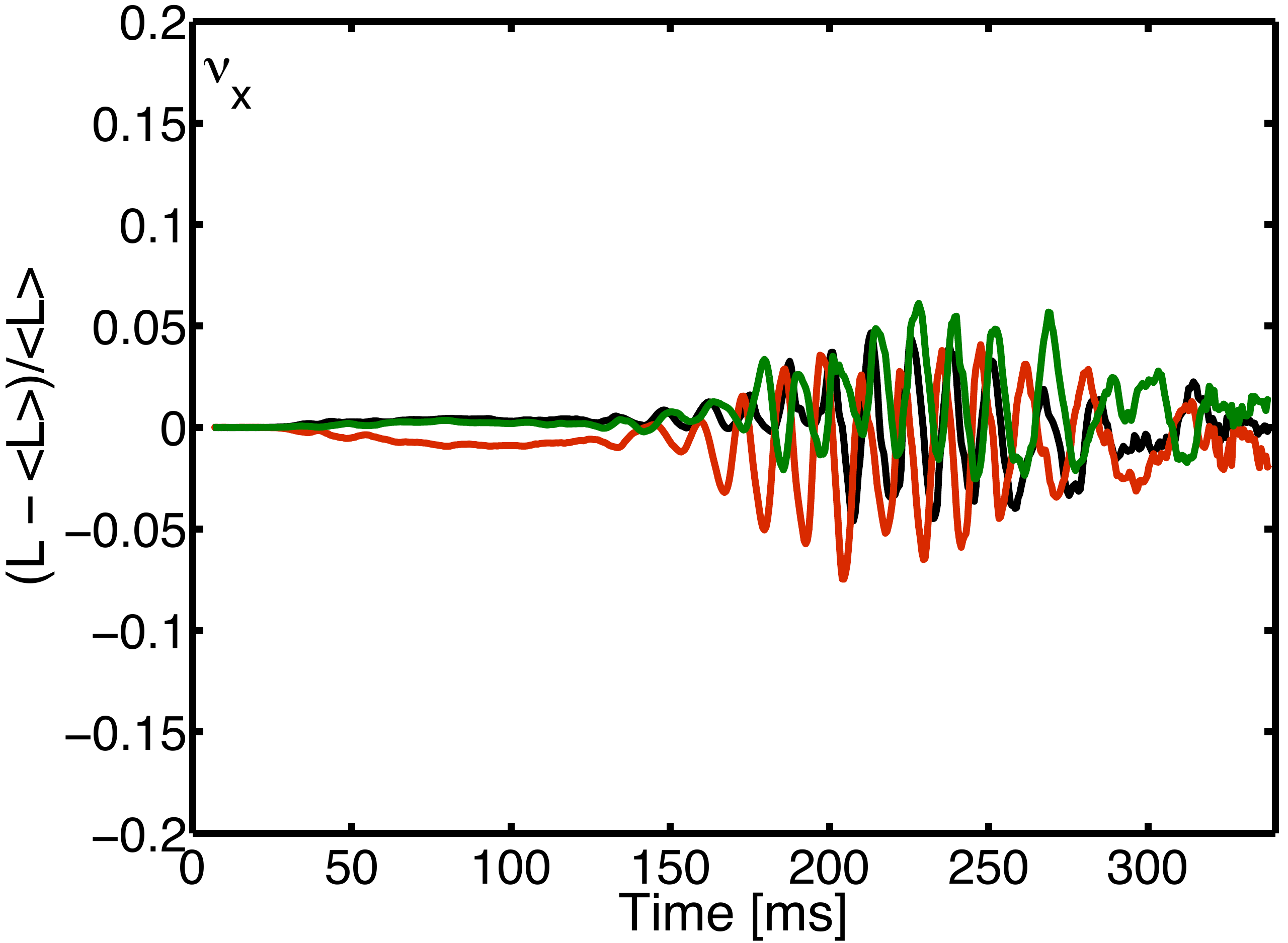}
\caption{Variation of the luminosity for $\nu_e, \bar{\nu}_e$ and
$\nu_x$ relative to the one computed as average over all directions
for the $20\ M_\odot$ SN progenitor at 10~kpc in analogy to Fig.~\ref{fig:lum11}.
The green and orange curves refer to locations of the observer close to the plane
where SASI develops and on opposite sides of the emitting sphere. The
black line refers to a location of the observer far from the SASI plane where the
modulation of the neutrino signal due to SASI is smaller.
\label{fig:lum20}}
\end{figure}

In contrast to the $11.2\,M_\odot$ case, the 20 and $27\,M_\odot$ progenitors
show large-amplitude modulation of the neutrino signal due to SASI spiral
modes, which cause accretion variations and corresponding fluctuations
of the neutrino emission.
The LESA phenomenon also occurs for these progenitors. Even though
LESA persists during the phases of violent sloshing of the shock-wave radius,
it is somewhat masked during the SASI episodes,  as explained in our LESA
paper~\cite{Tamborra:2014aua}. We focus first on the $27\,M_\odot$
progenitor to facilitate comparison with the previous discussions of this
model~\cite{Hanke:2013ena,Tamborra:2013laa}.

Figure~\ref{fig:lum27} shows the luminosity evolution, relative to the
directional average, for the three flavors in analogy to
Fig.~\ref{fig:lum11}. However, here we do not use the LESA axis and locate
the observers in directions where the SASI amplitude is particularly large
during the first SASI episode (light blue and violet lines) and a third
direction where it is small (black).  The SASI-implied modulations, on the
other hand, are such that $L_{\nu_e}$ and $L_{\bar\nu_e}$ vary in phase with
each other (see Fig.~\ref{fig:violet27}). The SASI variation of the neutrino
signal is up to 15\% for $\nu_e$, even larger for $\bar{\nu}_e$, and still
around $5\%$ for~$\nu_x$.

While both SASI and convection can lead to large-scale shock deformations,
SASI is distinguished by a characteristic quasi-periodic oscillatory nature.
As discussed in Ref.~\cite{Hanke:2013ena}, the SASI sloshing axis initially
wanders and then stabilizes as the sloshing of the shock further grows in
amplitude and violence. When SASI starts to grow vigorously, predominantly
sloshing occurs, whereas later a transition to a spiral mode takes place,
associated with a circular motion of the maximum shock radius. Both
SASI and convective regimes are easily recognized in Fig.~\ref{fig:lum27}.
For 120--260~ms, SASI sloshing and spiral modes occur, for 260--410~ms
convection dominates, and then a second SASI episode takes place up to the
end of our simulation (cf.~Figs.~1, 2 and 6 of Ref.~\cite{Hanke:2013ena}).

The plane where spiral motions develop remains relatively stable until the
maximum amplitude is reached and SASI dies down. During the first SASI spiral
phase, the  plane where it develops is roughly perpendicular to the vector
$\mathbf{n}=(-0.35,0.93,0.11)$ in the SN simulation grid, i.e., there is no
alignment with the axis of the spherical polar grid~\cite{Hanke:2013ena}. The
second SASI phase develops in a plane different from the first
one. Therefore, the three fixed directions shown in Fig.~\ref{fig:lum27} are
no longer optimal relative to a maximum SASI effect.

In Fig.~\ref{fig:violet27} the luminosity evolution, relative to the
directional average, is shown as a function of time for the three flavors,
along the direction plotted in violet in Fig.~\ref{fig:lum27}. The LESA
phenomenon, while somewhat masked during the SASI episodes, clearly appears
during the convective phase between the SASI episodes, in the form of a
hemispheric asymmetry between $\nu_e$ and $\bar\nu_e$ luminosity.  The
relative LESA amplitude of $\bar{\nu}_e$ is opposite in sign to the one of
$\nu_e$ and the amplitude for $\nu_x$ is smaller, with its sign
correlated with the one of $\bar\nu_e$. The SASI modulations, on
the other hand, have the same sign for all flavors, but a smaller amplitude
for $\nu_x$.

In order to discuss the directional dependence of the  SASI modulation of the
neutrino signal, Fig.~\ref{fig:goodbad_dir} shows the neutrino flux
properties (luminosity, mean energy and shape parameter $\alpha$) for the
three flavors along the same three directions, respectively corresponding to
the violet, black, and light blue curves in Fig.~\ref{fig:lum27} and named by
color. Although the neutrino flux properties are similar for all directions,
the modulation of the signal along ``Black'' and ``Light Blue'' directions is
much less pronounced during the first SASI episode than the modulation along
the ``Violet'' direction.

Figure~\ref{fig:lum27times} shows sky maps of the relative luminosity of
$\bar{\nu}_e$ for $t=217$, 225 and 230~ms, i.e., corresponding to subsequent
SASI maximum and minimum signal amplitudes. Comparing the three snapshots
there is again a total variation of $\sim 20\%$ for the different angular
positions. Looking at the hottest and coldest spot in the three time slices,
it is clear how the SASI sloshing motions proceed.

We repeat the same analysis as before for the $20\ M_\odot$ progenitor.
Figure~\ref{fig:lum20}, in analogy to Fig.~\ref{fig:lum27}, shows  the
relative luminosity. This progenitor exhibits only one SASI episode for $t
\ge 160$~ms, lasting for a longer time than for the $27\ M_\odot$ progenitor.
\begin{figure}
\centering
\includegraphics[width=0.9\columnwidth]{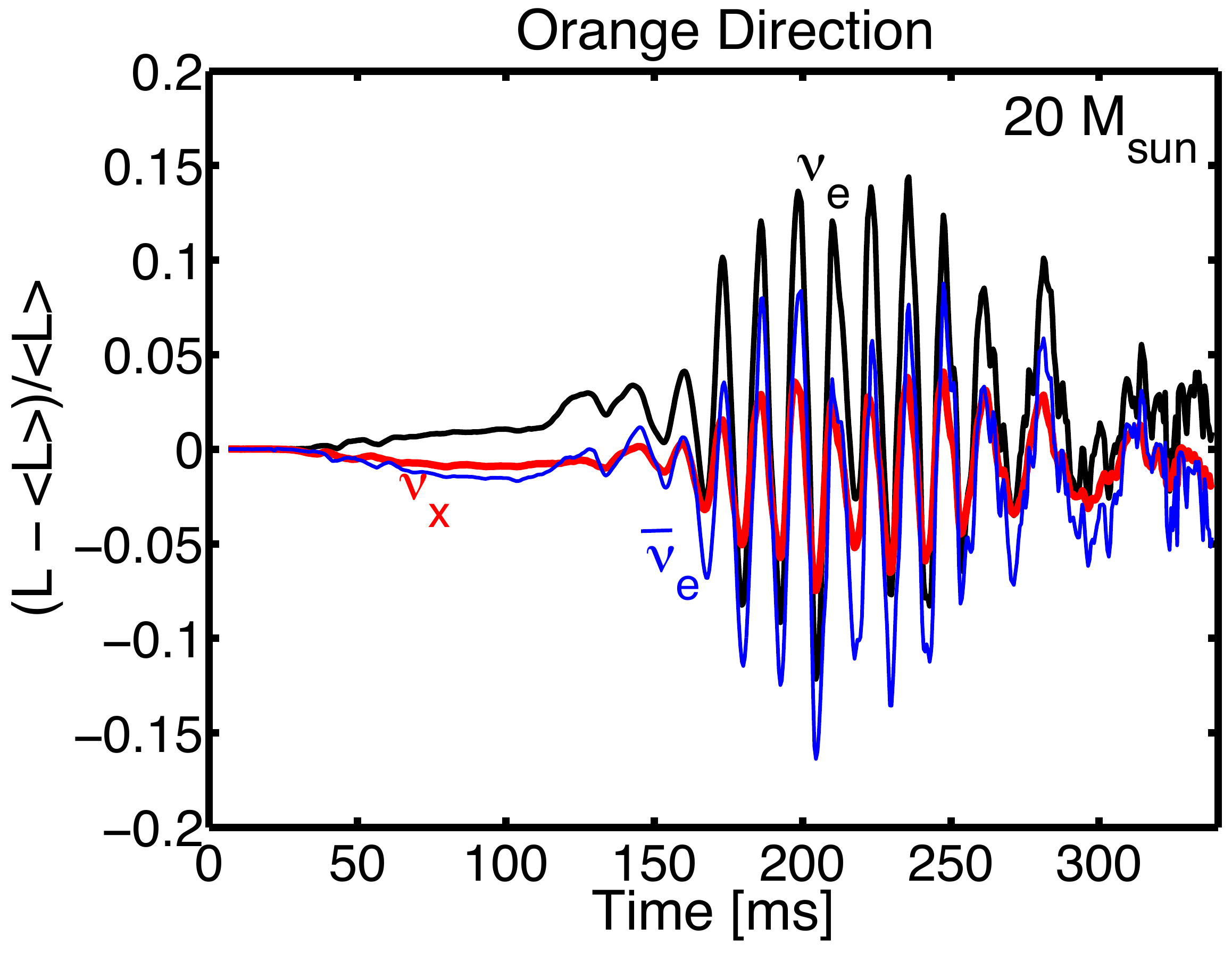}\\
\caption{Luminosity evolution relative to $\langle L \rangle$ for the $20\,M_\odot$ progenitor
for the three species, as seen by a distant observer
along the ``Orange'' direction, corresponding to the curve of this color in  Fig.~\ref{fig:lum20}.
See Fig.~\ref{fig:violet27} for comparison.
\label{fig:orange20}}
\end{figure}
\begin{figure}
\centering
\includegraphics[width=0.9\columnwidth]{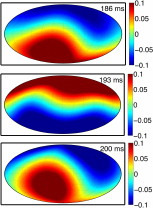}
\caption{Sky maps of $L_{\bar\nu_e}$ relative to the $4\pi$ average
for the $20\ M_\odot$ SN progenitor at $t=186$, 193 and 200~ms,
corresponding to subsequent SASI maxima and minima in analogy to Fig.~\ref{fig:lum27times}.
\label{fig:lum20times}}
\end{figure}
The SASI-implied modulations are again such that $L_{\nu_e}$ and
$L_{\bar\nu_e}$ vary in phase with each other, as clearly visible in
Fig.~\ref{fig:orange20}.  Traces of LESA appear in the hemispheric
asymmetry between the $\nu_e$ and $\bar\nu_e$ luminosity, especially
before SASI sets in ($t < 160$~ms) when the relative variation of
$\bar{\nu}_e$ is opposite in sign to the one of $\nu_e$. We find
maximum fluctuations of the signal of 17\% and minimum of
7\%. Figure~\ref{fig:lum20times}, similar to
Fig.~\ref{fig:lum27times}, shows sky maps of the relative
$\bar{\nu}_e$ luminosity for three snapshots ($t=186$, 193 and 200 ms)
corresponding to the SASI maximum and minimum signal amplitudes. As
for the $27\ M_\odot$ SN progenitor, a total variation of $\sim 20\%$
for the different angular positions occurs. The SASI spiral mode
develops in a plane perpendicular to $\mathbf{n}=(-0.56,-0.81,-0.20)$
in the SN simulation grid, i.e., in a different plane than in the
$27\,M_\odot$ case, as evident from a comparison of
Figs.~\ref{fig:lum27times} and~\ref{fig:lum20times}. In fact the
SASI plane is randomly selected and bears no relation to the numerical
grid in the case of a non-rotating model.

\subsection{The LESA phenomenon in presence of SASI}

\begin{figure*}
\centering
\includegraphics[width=0.4\textwidth]{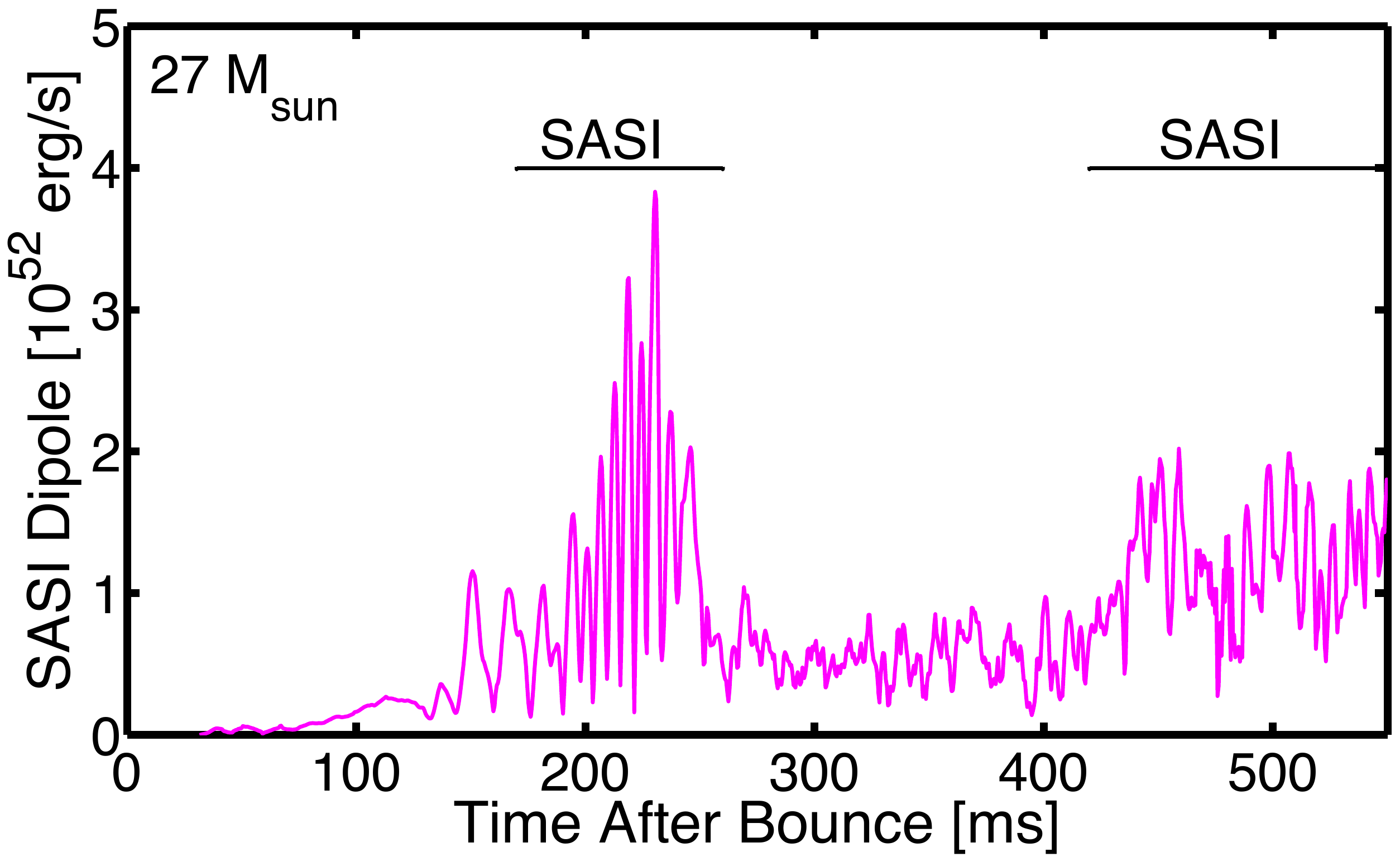}
\includegraphics[width=0.4\textwidth]{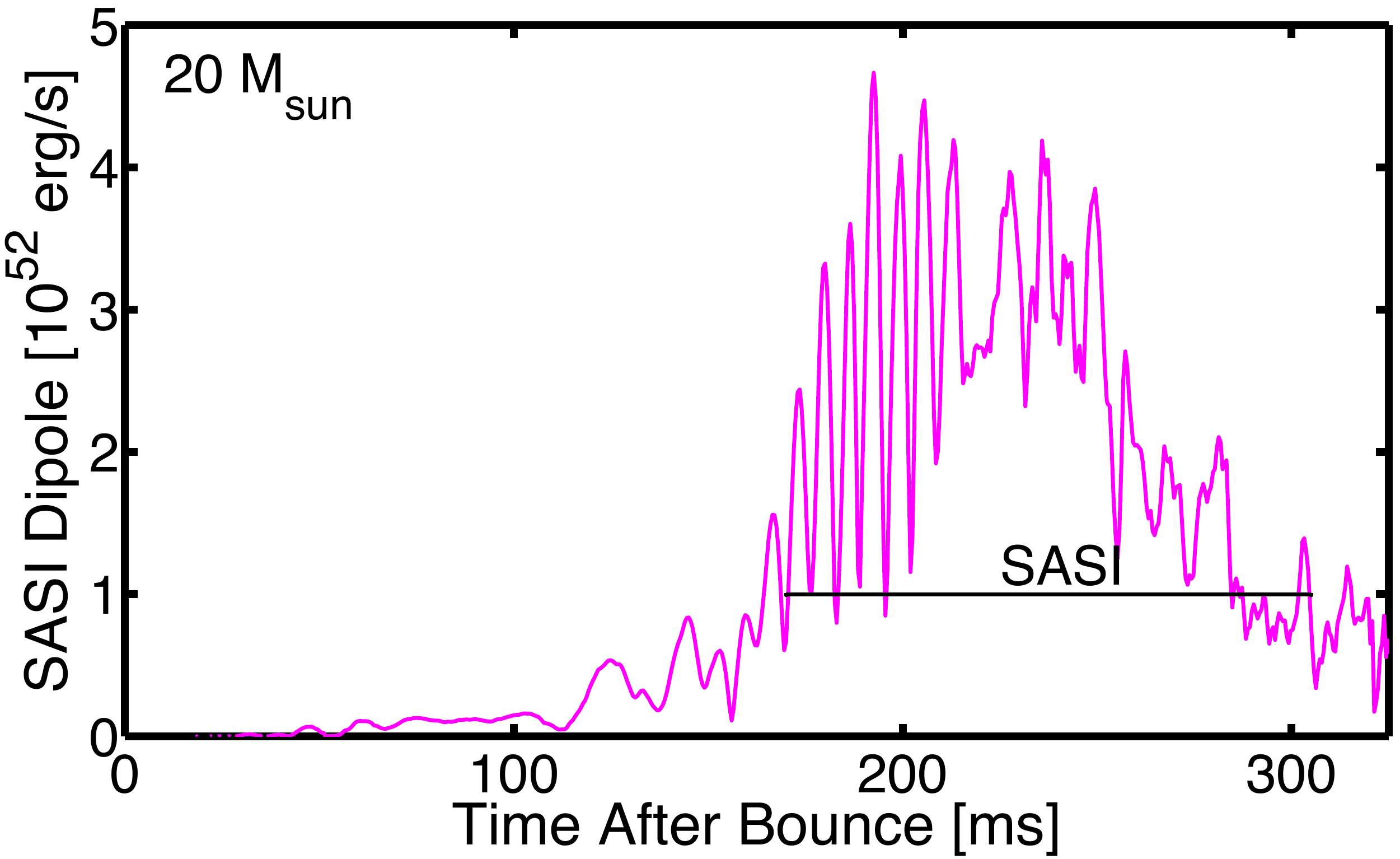}\\
\includegraphics[width=0.4\textwidth]{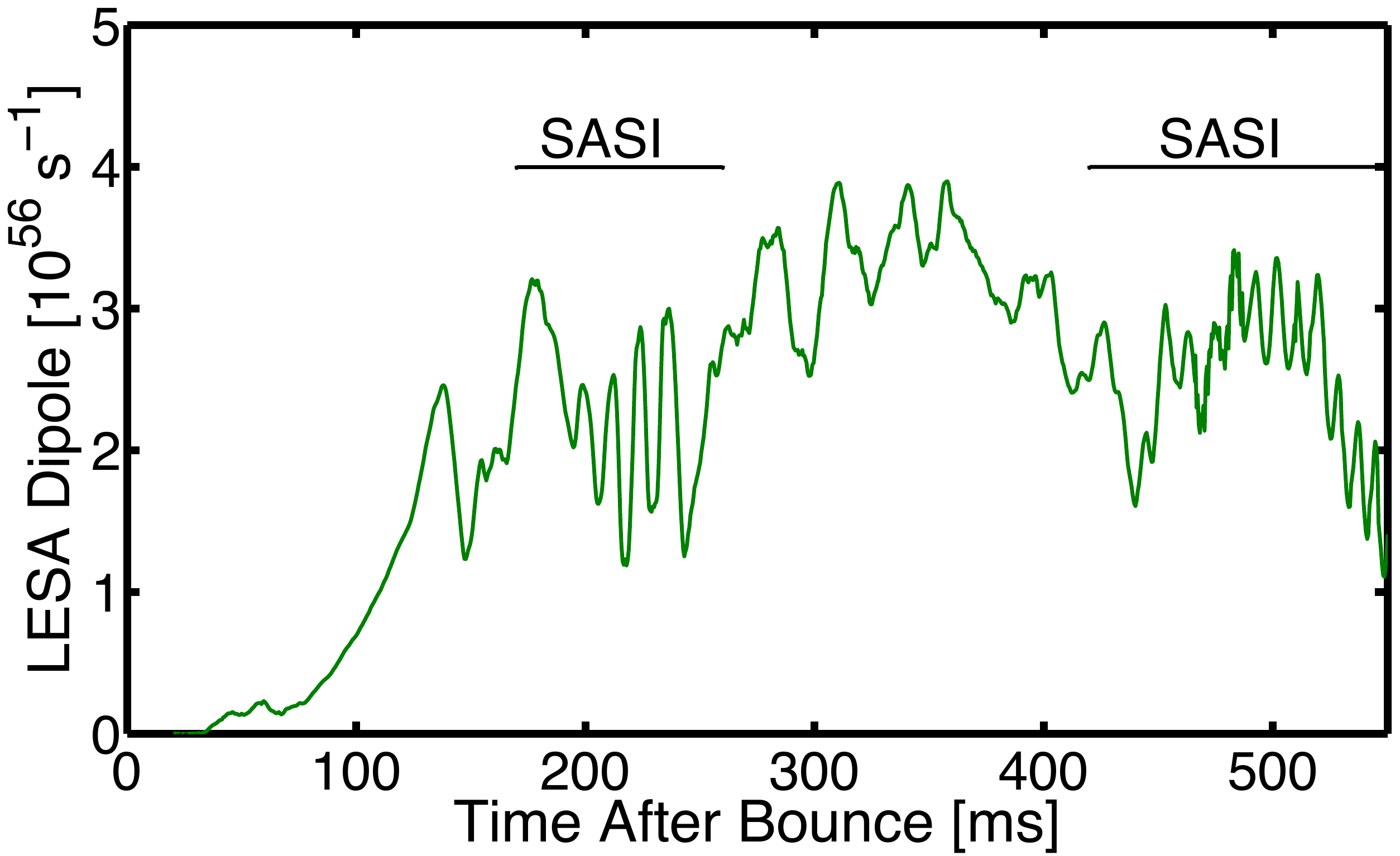}
\includegraphics[width=0.4\textwidth]{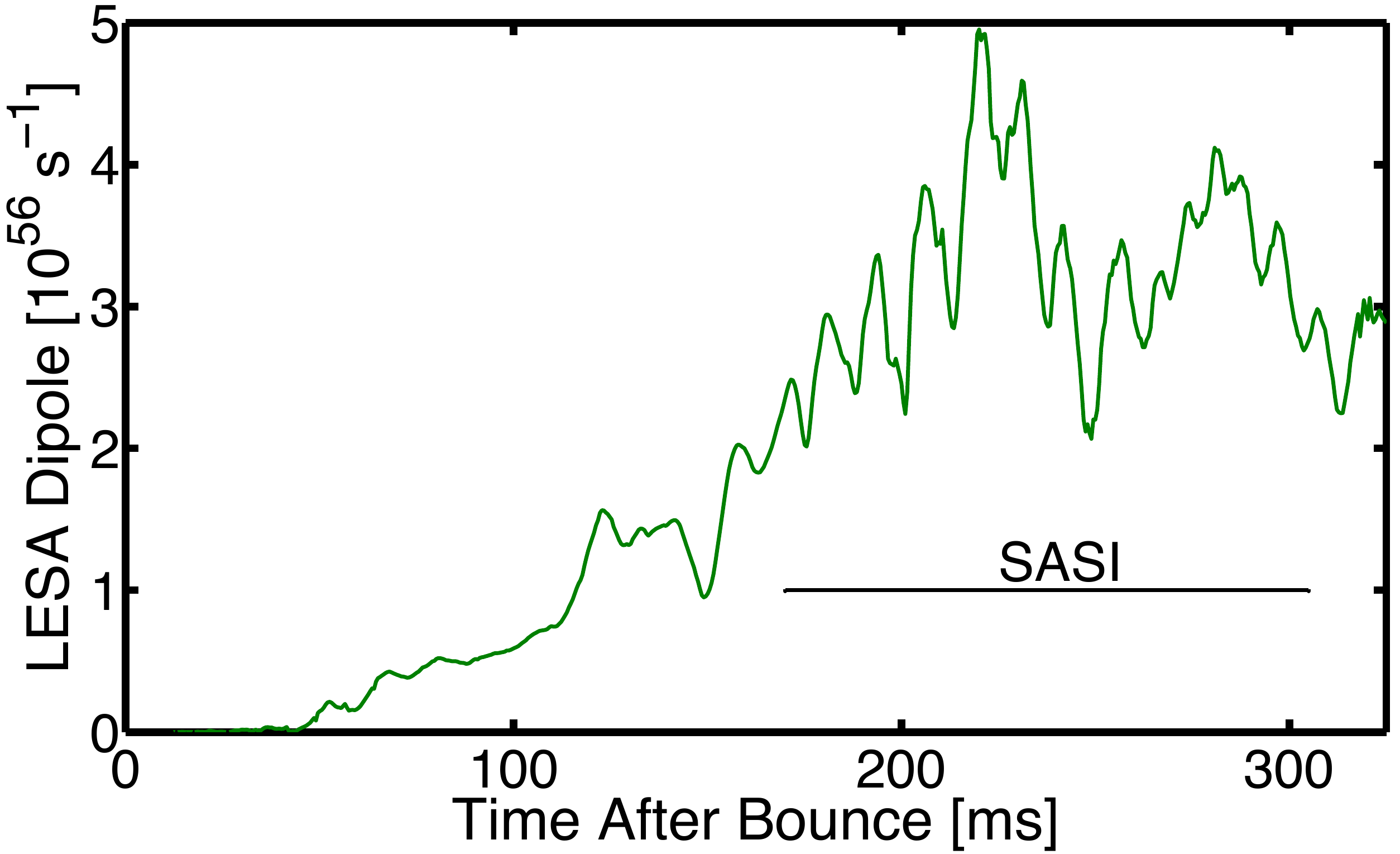}
\caption{SASI vs.\ LESA dipole as a function of time for the 
 $27\ M_\odot$ (left) and the $20\ M_\odot$ (right) simulations. We
 describe the SASI dipole (top panels) in terms of the neutrino energy flux of all six
 neutrino species. For the $27\ M_\odot$ model (left), it
 reaches a maximum of 0.16 relative to its monopole at 200~ms, 
 while for the the $20\ M_\odot$ model
 (right) the maximum is reached at 180~ms with the same relative strength.
 The LESA dipole (bottom panels) is described in
 terms of the lepton-number flux ($\nu_e$ minus $\bar\nu_e$) and
 reaches a relavive maximum of 2 times the monopole at 500~ms for the
 $27\ M_\odot$ case (left) and of 1.4 times the monopole at 280~ms for
 $20\ M_\odot$.
\label{fig:LESAdip_SASIdip}}
\end{figure*}
\begin{figure*}
\centering
\includegraphics[width=0.45\textwidth]{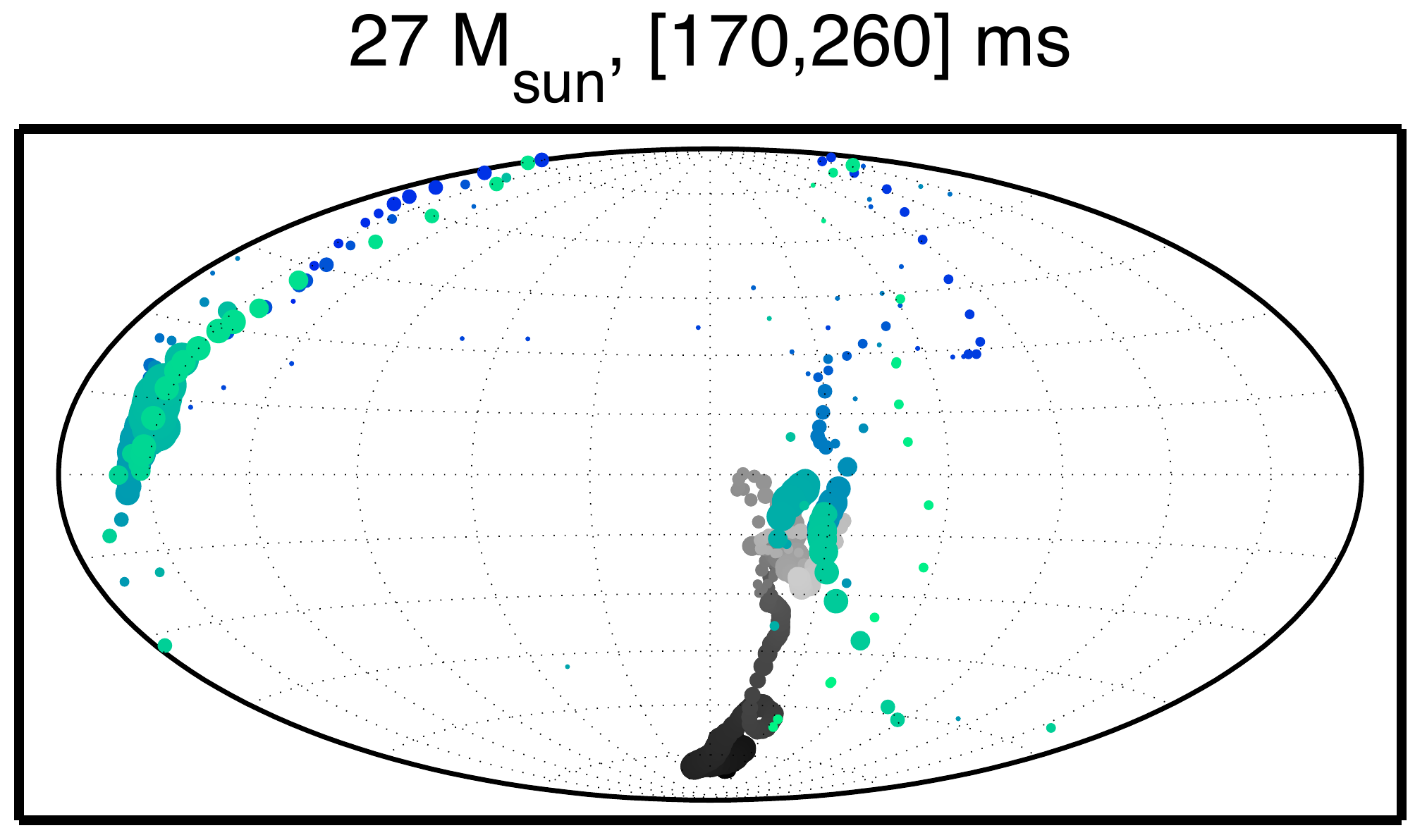}
\includegraphics[width=0.45\textwidth]{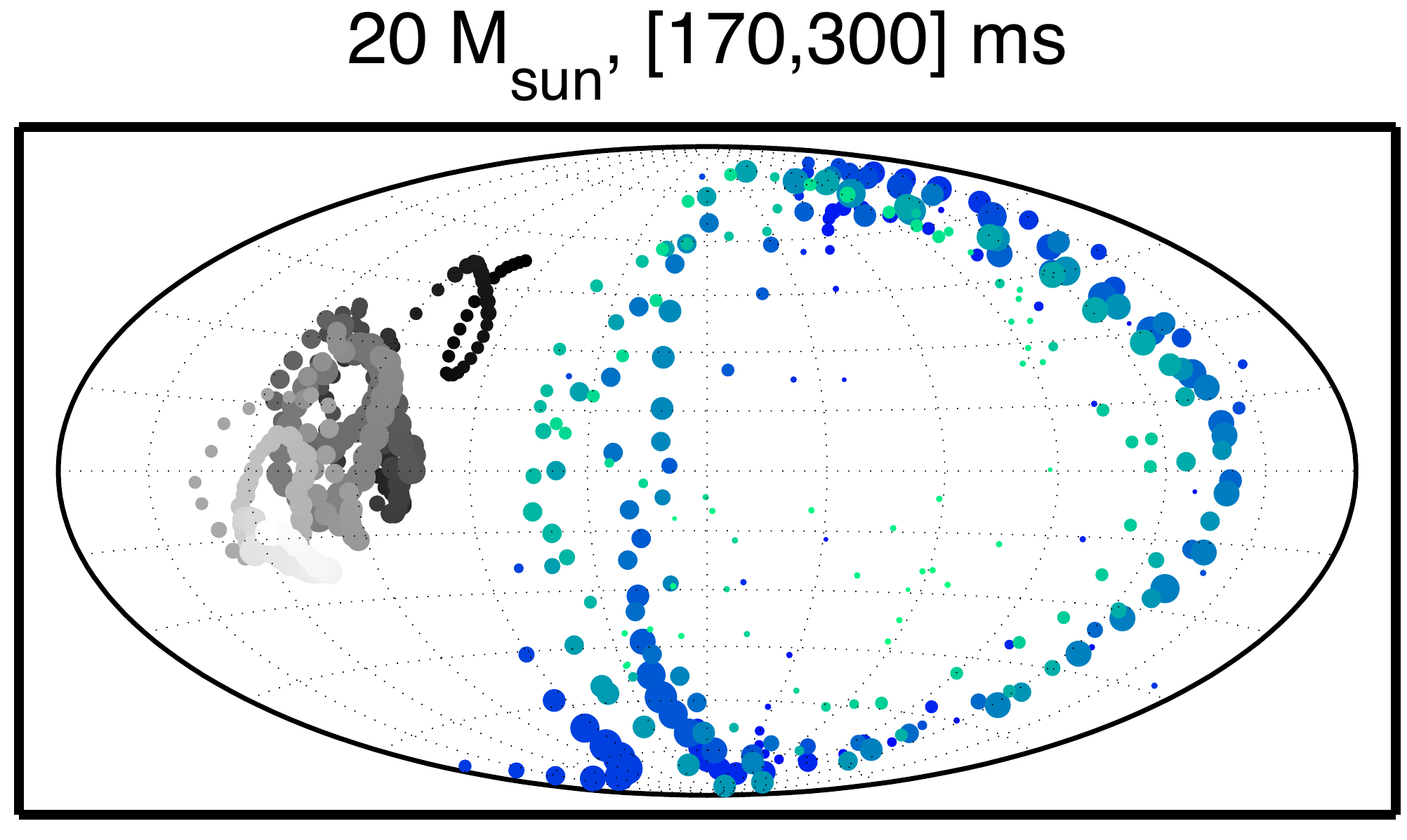}
\caption{Sky maps showing the evolution of the LESA direction
(i.e., direction of the dipole of the lepton number flux) in gray
color scale and of the SASI direction (i.e., direction of the dipole
of the neutrino energy flux of all six neutrino species) in blue color
scale for the $27\ M_\odot$ SN progenitor (left) and the $20\
M_\odot$ SN progenitor (right) during the periods of strongest
SASI activity.  The color hues become lighter as time progresses. The
size of the dots scales with the length of the plotted dipole
vector. While for the $27\ M_\odot$ SN progenitor the LESA dipole is
close to the SASI plane, for the $20\ M_\odot$ case they are nearly
perpendicular to each other and no clear correlation exists between
the LESA dipole direction and the SASI one.
\label{fig:LESA_SASI}}
\end{figure*}

The LESA phenomenon is characterized by a maximum of the $\nu_e$
emission coincident with a minimum of the $\bar{\nu}_e$ emission
(i.e., the amplitude of the $\nu_e$ emission is anti-correlated with
the amplitude of the $\bar{\nu}_e$ emission), whereas SASI is
responsible for correlating the amplitude variations of the $\nu_e$
and $\bar{\nu}_e$ signals (see
Figs.~\ref{fig:lum27}, \ref{fig:violet27}, \ref{fig:lum20},
and \ref{fig:orange20}).  In order to investigate in greater 
detail the
LESA phenomenon in the presence of SASI for our two heavier
progenitors, we consider the LESA dipole, i.e., the dipole component of
the lepton number flux ($\nu_e$ minus $\bar{\nu}_e$), following the
definition adopted in Sec.~3.1 of Ref.~\cite{Tamborra:2014aua}
(see also their Fig.~3), and the SASI dipole,
i.e., the dipole component of the neutrino energy flux of all flavors
($\nu_e + \bar{\nu}_e + 4\ \nu_x$). Note that we choose the total
neutrino energy flux to define the SASI dipole instead of the total
number flux because SASI modulates the total energy flow, including
the mean neutrino energies~\cite{Tamborra:2013laa}. The monopole
component of the total neutrino energy flux of all flavors corresponds
to the sum of luminosities for all flavors, as shown in
Figs.~\ref{fig:averageSN} (top panel) and \ref{fig:luminosities20} for
the $27\ M_{\odot}$ and $20\ M_\odot$ SN progenitors, respectively.
   
Figure~\ref{fig:LESAdip_SASIdip} shows the length of the dipole vector
of the neutrino energy flux of all six neutrino species (top panels)
and of the dipole component of the lepton number flux (bottom panels)
for the $27\ M_\odot$ (left) and for the $20\ M_\odot$ (right)
progenitors as a function of time.  Note that the LESA dipole is
different from zero, even during the SASI
episodes~\cite{Tamborra:2014aua}.  This means that the LESA mechanism
is active during SASI episodes, even if the LESA dipolar behavior
is not clearly visible in the neutrino signal in
Figs.~\ref{fig:lum27}, \ref{fig:violet27}, \ref{fig:lum20},
and \ref{fig:orange20} because it is masked by strong SASI modulations.

The ``SASI dipole'' of the $27\ M_\odot$ progenitor does
not completely vanish between SASI episodes because large-scale
convection also causes an overall emission dipole. During the first
SASI episode it increases strongly, reaching its maximum around
$200$~ms where it is about 16\% of its monopole (total energy flux,
i.e., the sum of the luminosities as plotted on the top panel of
Fig.~\ref{fig:averageSN}). During the second SASI phase it is at most
10\% of its monopole. On the other hand, the LESA dipole quickly grows
up to 150~ms. The LESA dipole is almost two times the monopole at
about 500~ms. The SASI activity of the $20\ M_{\odot}$ model is more pronounced
compared to the the $27\ M_{\odot}$ simulation and the SASI dipole
is correspondingly larger. However, in relative terms it also reaches
a maximum of 16\% of the monopole strength at 180~ms and then decreases.
On the other hand, the
ratio between the LESA dipole and monopole is maximum at $280$~ms and
is about 1.4. It is interesting to notice that the general trend as
a function of time is the same between LESA and SASI dipoles, but it
is strongly progenitor dependent. In particular the dipole grows
during the SASI activity for the $20\ M_\odot$ SN progenitor, while it
is on average stationary during SASI for the $27\ M_\odot$ SN
progenitor.

Figure~\ref{fig:LESA_SASI} shows the track of the LESA dipole in gray
and the SASI dipole in blue hues during the SASI episodes for the $27\ M_\odot$ 
(left) and for the $20\ M_\odot$ (right) SN progenitors, in
order to investigate a possible correlation between the 
LESA dipole and the plane of the SASI sloshing and spiral modes. 
While
for the $27\ M_\odot$ case, the SASI spiraling drives the LESA dipole
to wander in the SASI plane, this does not happen for the $20\
M_\odot$ SN progenitor. The neutrino SASI dipole trajectory closely
reproduces the shock-deformation trajectory shown in Fig.~8 of
Ref.~\cite{Tamborra:2014aua}.
 
The LESA dipole direction as well as the SASI dipoles are progenitor
dependent (see also Fig.~3 of Ref.~\cite{Tamborra:2014aua}), and none
of them are correlated with the numerical grid of the simulation.
The mutual interaction between SASI and LESA seems to be strongly
progenitor dependent, although from this preliminary analysis it is
clear that these are two separate phenomena. We here refrain from
drawing any firm conclusion on the interaction between LESA and SASI
since a much deeper understanding of the LESA phenomenon and its
origin is required and hydrodynamical simulations of more SN progenitors 
are needed to favor a better understanding of the coexistence between 
the two phenomena.

\section{Flavor oscillations}
\label{sec:oscillations}

Neutrino transport in SN models is treated in the weak-interaction basis of
flavors. In our three-species treatment, we use $\nu_e$, $\bar\nu_e$ and $\nu_x$,
neglecting weak-magnetism effects that distinguish between neutral-current
scattering of $\nu_\mu$ ($\nu_\tau$) and $\bar\nu_\mu$ ($\bar\nu_\tau$). We also ignore
the possible presence of muons that would allow charged-current processes for
$\nu_\mu$ and $\bar\nu_\mu$ in the deep interior of the proto-neutron star. Most importantly, we ignore
flavor conversion caused by flavor mixing. The justification for this
simplification is the strong matter effect that effectively ``de-mixes''
neutrinos, i.e., the propagation eigenstates essentially coincide with the
weak-interaction eigenstates~\cite{Wolfenstein:1977ue}.

However, as neutrinos stream away from the SN core, the matter effect
decreases and eventually flavor conversion becomes important. What is
measured in a detector crucially depends on neutrino flavor oscillations along the way.

In the simplest traditional picture, the slowly-varying matter profile
provides for adiabatic flavor conversion, the so-called Mikheev-Smirnov-Wolfenstein
(MSW) effect~\cite{Wolfenstein:1977ue, wolf}. In particular, the
recent measurement of the third mixing angle
$\sin^2(2\Theta_{13})=0.095\pm0.010$ \cite{Beringer:1900zz}, being fairly
large, implies that the entire three-flavor conversion process would indeed be
adiabatic~\cite{Dighe:1999bi}. For the normal neutrino mass hierarchy (NH), the
$\bar\nu_e$ survival probability is $\bar p_{\rm
NH}=\cos^2\Theta_{12}\sim0.70$, whereas for the inverted ordering (IH) it is $\bar
p_{\rm IH}=0$ \cite{Dighe:1999bi}. Therefore, a detector measuring
$\bar\nu_e$ by inverse beta-decay (IBD) will see in NH a superposition of roughly 70\% of the
original $\bar\nu_e$ flux spectrum with 30\% of the $\bar\nu_x$ flux
spectrum, whereas in IH it will detect the original $\bar\nu_x$ flux
spectrum at the source.

This simple prediction can get strongly modified by two effects. The density
profile can be noisy and show significant stochastic
fluctuations~\cite{Kifonidis:2003fv,Kifonidis:2005yj,Scheck:2006rw,Hammer:2009cn,Muller:2011yi} that can
modify the adiabatic conversion~\cite{Loreti:1995ae,Fogli:2006xy,Friedland:2006ta,Kneller:2010sc,Lund:2013uta,Borriello:2013tha}.
Such effects would be especially expected in
the turbulent medium behind the shock wave, i.e., the relevance pertains in
particular to neutrino propagation after the explosion has set in and the
shock wave travels outward. However, we are here concerned with the
standing-shock phase and flavor conversion outside of the shock-wave radius.

Of greater importance is then the impact of neutrino-neutrino refraction which can
lead to self-induced flavor conversion, usually at a smaller radius than the
MSW effect~\cite{Duan:2010bg}. It can put the MSW result effectively upside
down and can lead to novel spectral features (spectral splits)
\cite{Duan:2006an, Fogli:2007bk, Raffelt:2007cb, Fogli:2008pt, Dasgupta:2009mg, Fogli:2009rd, Dasgupta:2010cd}. On the
other hand, self-induced flavor conversion can be suppressed by the
``multi-angle matter effect'' \cite{EstebanPretel:2008ni} and this may be
typical in many cases, re-instating the traditional
scenario~\cite{Chakraborty:2011nf, Sarikas:2011am, Saviano:2012yh}. On the
other hand, what exactly happens when self-induced conversion is not
suppressed remains poorly understood because of a number of complications
that have only recently been appreciated~\cite{Cherry:2012zw, Sarikas:2012vb,
Raffelt:2013rqa, Raffelt:2013isa, Hansen:2014paa, Mirizzi:2013rla,
Mirizzi:2013wda, Chakraborty:2014nma, Mangano:2014zda}. In addition, the
direction-dependent neutrino flux properties and especially the LESA
phenomenon throw in additional uncertainties that have not been studied yet.

In this situation we can but state that the $\bar\nu_e$ flux arriving at the
detector will be some superposition, possibly depending on energy, of the
original $\bar\nu_e$ and $\bar\nu_x$ flux spectra. We therefore consider two
extreme cases. One is that the detector measures the original $\bar\nu_e$
flux, the other assumes a complete flavor swap and the detector measures what
was the $\bar\nu_x$ flux at the source.

\section{Detection of Signal Modulations}
\label{sec:detector}

\subsection{Detector Models}

Detecting the SASI-imprinted modulations in the high-statistics neutrino
signal of the next galactic SN would go a long way in studying SN
hydrodynamics. What are the opportunities for such a detection?

In the largest operating detector, IceCube, and the future Hyper-Kamiokande,
neutrinos are primarily detected by IBD, $\bar\nu_e+p\to
n+e^+$, through the Cherenkov radiation of the final-state positron. We will
ignore the small additional contribution from elastic scattering on
electrons.  The signature for fast time variations is limited by random
fluctuations (shot noise) of the detected neutrino time sequence.

In IceCube \cite{Abbasi:2011ss}, usually at most one Cherenkov photon from a
given positron is detected, i.e., every measured photon signals the arrival
time of a neutrino and in this sense provides superior signal statistics. In
rare cases, two or more photons from a single neutrino are detected,
depending on neutrino energy, allowing one to extract interesting spectral
information from time-correlated photons \cite{Demiroers:2011am}, but this
intriguing effect is not of direct interest here. The instantaneous signal
count rate caused by IBD in a single optical module (OM)
is~\cite{Abbasi:2011ss}
\begin{equation}\label{eq:rate-single}
r_{\rm IBD}=n_p \int dE_e\int dE_\nu\,F_{\bar{\nu}_e}(E_{\nu})\,
\sigma'(E_e, E_\nu)\,N_{\gamma}(E_e)\,V_{\gamma}^{\rm eff},
\end{equation}
where $n_p= 6.18 \times 10^{22}~{\rm cm}^{-3}$ is the number density of
protons in ice (density $0.924~{\rm g}~{\rm cm}^{-3}$), $E_e$ is the
final-state positron energy, $V_{\gamma}^{\rm eff} = 0.163 \times 10^{6}~{\rm
cm}^{3}$ the average effective volume for a single photon detection,
$N_{\gamma}(E_e) = 178\,E_e/{\rm MeV}$ is the energy-dependent number of
Cherenkov photons, and $\sigma'(E_e, E_\nu)=d\sigma(E_e, E_\nu)/dE_e$ is the
IBD cross section, differential with respect to the positron energy. 

We correct the positron energy, $E_e\to E_e+1~{\rm MeV}$, because gamma rays
from positron annihilation and neutron capture produce additional recorded
energy \cite{Abbasi:2011ss}. Moreover, the IceCube rate from IBD is about
94\% of the total, so we apply a correction factor
\begin{equation}\label{eq:rate-single2}
r=r_{\rm IBD}/0.94
\end{equation}
to account approximately for all channels.

Every OM shows a background rate of around 540~Hz, including correlated
events. Introducing an artificial dead time of $t_{\rm dead}=250~\mu{\rm s}$
after every hit reduces the background to a single rate of about 286~Hz at
the cost of about 13\% dead time. More specifically, the signal reduction by
this dead-time effect is $0.87/(1+r\,t_{\rm dead})$. Therefore, the overall
SN signal rate is
\begin{equation}\label{eq:rate-full}
R_{\rm IC}=N_{\rm OM}\,\frac{0.87\,r}{1+r\,t_{\rm dead}}\,,
\end{equation}
where $N_{\rm OM}=5160$ is the number of OMs in IceCube.

In previous studies of the IceCube potential for detecting fast signal
variations \cite{Lund:2010kh,Lund:2012vm}, these various corrections had not
been included. Moreover, a simple approximate expression for the IBD cross
section was used. As in our companion {\it Physical Review
Letter}~\cite{Tamborra:2013laa}, we here use the IBD cross section provided
in Ref.~\cite{Strumia:2003zx}, which includes recoil, the
neutron-proton mass difference, the positron mass, and nucleon form factors.
If the $\bar\nu_e$ spectrum is described by a Gamma distribution (see
Appendix~\ref{sec:ibd}), the final-state positrons also follow such a
distribution with good approximation. In Appendix~\ref{sec:ibd} we give
analytic approximation formulas for the spectral parameters of the detected
positrons in terms of those of the primary $\bar\nu_e$.

For the example of our $27\,M_\odot$ model, Fig.~\ref{fig:averageSN} (bottom
panel) shows the single-OM IceCube rate $r$ as defined in
Eq.~(\ref{eq:rate-single2}) without dead-time effect. We show $r$ for
$\bar\nu_e$, ignoring flavor oscillations, and also for $\bar\nu_x$ under the
assumption of a full flavor swap $\bar\nu_e\leftrightarrow\bar\nu_x$. The
maximum rate is around 170~Hz, somewhat larger than half of the background
rate, so that $r\,t_{\rm dead}\sim 0.04$. In this case, dead-time effects
reduce the overall signal to about 84\% of the raw rate.

Incorporating dead time, the average single-OM background rate is 286~Hz.
After multiplying with $N_{\rm OM}=5160$ we find an overall background rate
of
\begin{equation}
R_{\mathrm{bkgd}} =1.48 \times 10^3\ \mathrm{ms}^{-1}\,.
\label{bkgd}
\end{equation}
For a SN at 10~kpc, this is about twice a typical signal rate. The
detectability of fast time variations is limited by random signal
fluctuations (shot noise) which originates from both the signal itself and
fluctuations of the background rate.

As for Hyper-Kamiokande~\cite{Abe:2011ts}, a next-generation megaton water
Cherenkov detector, we focus on the number of IBD events, expecting a
correction of a few percent due to the other neglected channels. The expected
rate is
\begin{equation}
R_{\rm HK}=N_p \int dE_e\int dE_\nu\,F_{\bar{\nu}_e}(E_{\nu})\,
\sigma'(E_e, E_\nu),
\label{eq:HKrate}
\end{equation}
where $N_p=4.96 \times 10^{34}$ is the number of protons for a
$0.74$~Mton Cherenkov detector~\cite{Abe:2011ts}. The advantage relative to
IceCube is that such a detector is essentially background free and, as a
plus, will provide event-by-event energy information. We found that although
the expected rate as function of time is almost three times lower than the
IceCube rate~\cite{Tamborra:2013laa}, the expected Hyper-Kamiokande rate has
the same modulation of the signal with slightly lower amplitude. We also
noticed (results not shown here) as convolving the expected signal rate with
powers of the energy, that the amplitude of the sinusoidal modulations is
enhanced.

\subsection{Detection Perspectives}

\begin{figure*}
\centering
\includegraphics[width=0.348\textwidth]{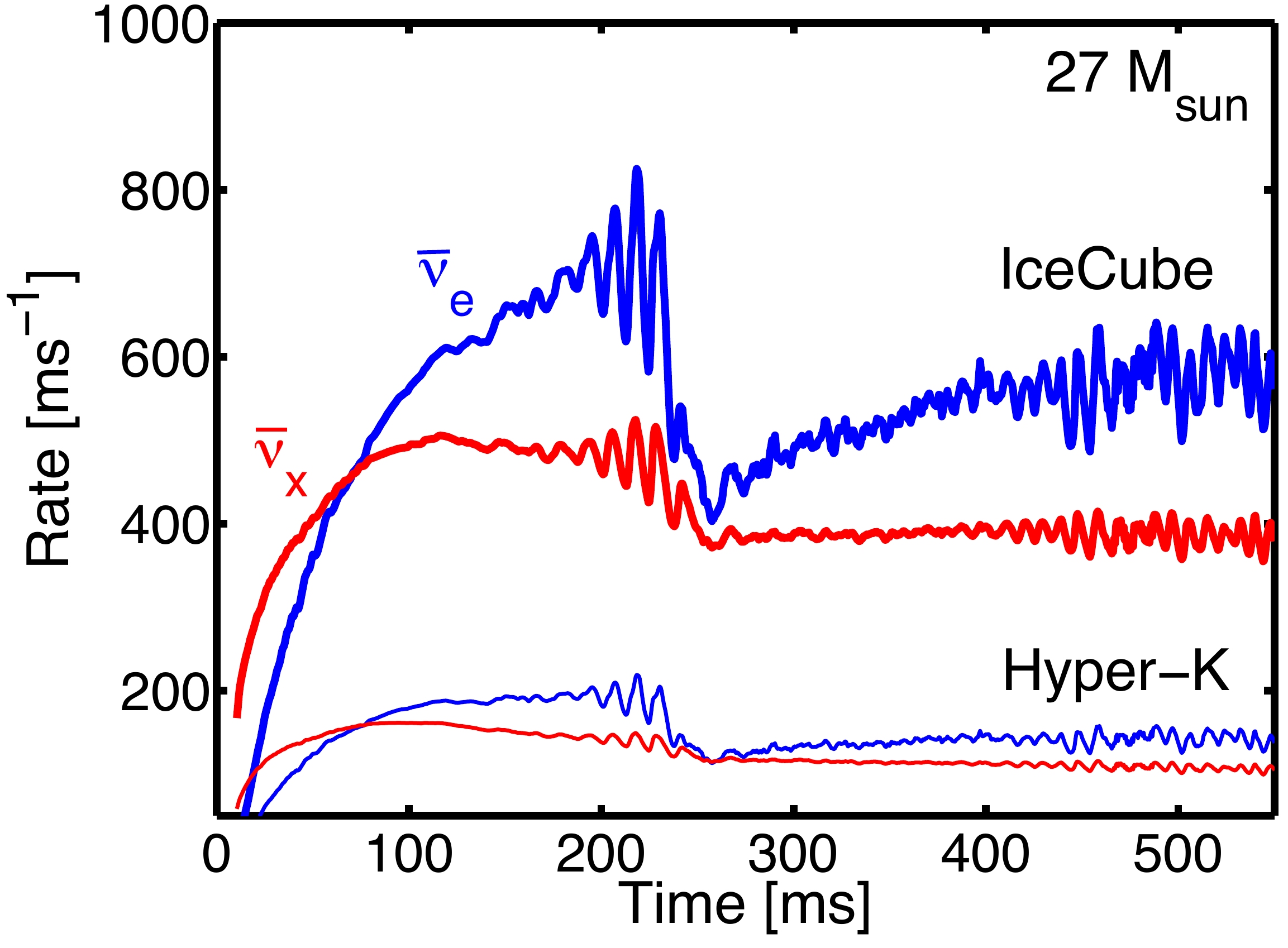}
\includegraphics[width=0.3\textwidth]{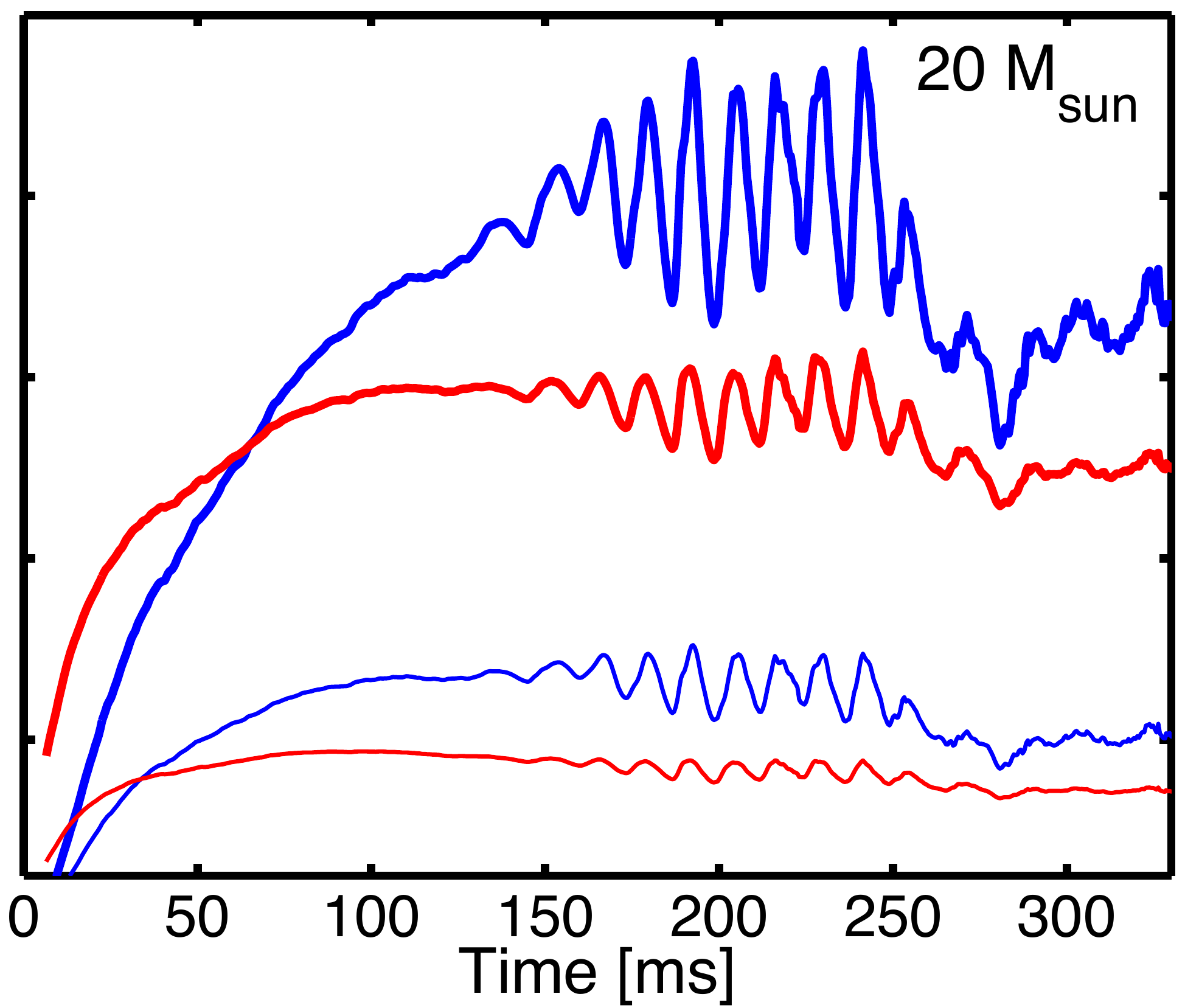}
\includegraphics[width=0.31\textwidth]{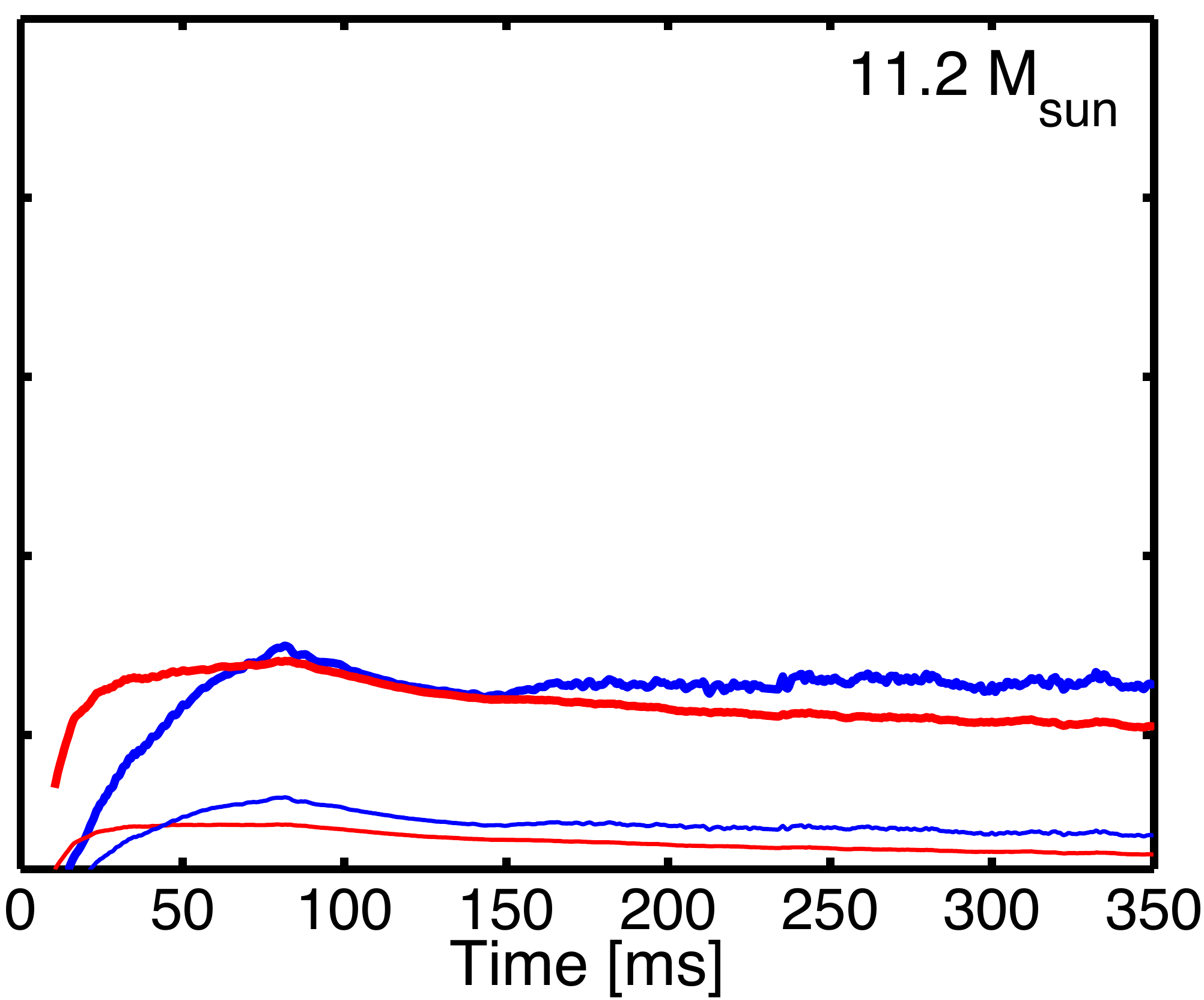}
\caption{IceCube and Hyper-Kamiokande detection rates $R$
 for our 27, 20 and $11.2\,M_\odot$ SN progenitors at a distance of 10~kpc. The
 rates are shown for $\bar\nu_e$ and for $\bar\nu_x$  (i.e., assuming full
 swap by flavor conversion). For the $27$ and $20\,M_\odot$ progenitors,
 the observer direction is close to the SASI plane where the signal modulation
 is strong.  \label{fig:ICHKrate}}
\end{figure*}

Figure~\ref{fig:ICHKrate} shows the expected IceCube and Hyper-Kamiokande
rates for 27, 20 and $11.2\ M_\odot$ SN progenitors, respectively, from left
to right, at a distance of 10~kpc and for $\bar{\nu}_e$ and $\nu_x$ (assuming
full swap by flavor conversion) fluxes. For the progenitors where SASI
develops (27 and $20\,M_\odot$ progenitors), we show the signal as seen by a
distant observer close to the SASI spiral plane where the signal modulations
are large. The IceCube rate was defined in Eq.~(\ref{eq:rate-full}) and the
Hyper-Kamiokande one in Eq.~(\ref{eq:HKrate}).

The relative amplitude of the SASI modulations is similar in the
$\bar{\nu}_e$ and $\bar{\nu}_x$ channels, although the $\bar\nu_x$ rate is
always lower that the $\bar{\nu}_e$ one. The origin of this effect is that,
although the luminosities show different amplitudes of SASI modulation
(Fig.~\ref{fig:luminosities}), the $\bar\nu_x$ spectrum is less pinched than
the $\bar\nu_e$ one (see $\alpha$'s for $\bar{\nu}_e$ and $\nu_x$ in
Fig.~\ref{fig:averageSN}). As discussed in our earlier
paper~\cite{Tamborra:2013laa}, in spite of shot noise, an observer located
along an optimal direction will be able to detect SASI modulations out to a
distance of 20~kpc (cf.\ Fig.~1 of Ref.~\cite{Tamborra:2013laa}). Note that
IceCube and Hyper-Kamiokande offer complementary information since IceCube
will be more suitable for SN at small distances where the shot noise is smaller.
Hyper-Kamiokande will be more useful at larger distances because it is
background free and the shot noise is dominated by fluctuations of the signal
itself~\cite{Tamborra:2013laa}.

As shown in Fig.~\ref{fig:lum27} (black line), SASI has not the same
intensity along all the directions and can be very weak. In order to
characterize where, on average, the modulations of the neutrino signal due to
SASI are stronger and therefore where an observer has more chances to detect
it, we define the following standard deviation of the IceCube rate for each
angular position of the observer~\cite{Tamborra:2013laa}
\begin{equation}
\sigma^2 = \int_{t_1}^{t_2} dt\ \left(\frac{R-\langle R \rangle}{\langle R
\rangle}\right)^2\ ,
\label{eq:sigma2}
\end{equation}
where $\left\langle R\right\rangle$ is the time-dependent average over
all directions.

\begin{figure}
\centering
\includegraphics[width=0.9\columnwidth]{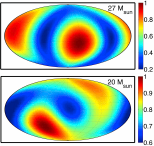}
\caption{Sky-map of $\sigma/\sigma_{\mathrm{max}}$ with
$\sigma^2$ defined in Eq.~(\ref{eq:sigma2}). Top panel: $27\,M_\odot$
progenitor during the first SASI phase (120--250~ms). Bottom panel:
$20\,M_\odot$ progenitor during 150--330~ms.
\label{fig:4pimap}}
\end{figure}

Figure~\ref{fig:4pimap} (upper panel) shows a sky-map of
$\sigma/\sigma_{\mathrm{max}}$ with $\sigma_{\mathrm{max}}$ the maximum of
$\sigma$ during the first SASI window of the $27\ M_\odot$ SN progenitor,
i.e., integrating over the time interval 120--250~ms. It is clear that the
SASI modulations are on average stronger, and will be experimentally
observable, for an extended region close to the SASI spiral plane, roughly
corresponding to 60\% of possible observer locations. The regions of strong
SASI modulations visible in Fig.~\ref{fig:4pimap} correspond to the hottest
and coldest regions in Fig.~\ref{fig:lum27times}. Of course, for several
SASI episodes or a strong drift of the main plane, some part of the SASI
activity may become visible in a larger fraction of all observer directions at different
times.

Figure~\ref{fig:ICHKrate} (middle  panel) shows the IceCube and
Hyper-Kamiokande rates for the $20\ M_\odot$ SN progenitor along one of the
directions where SASI modulation of the neutrino signal is strong. Only one
SASI phase occurs for this progenitor and SASI is somewhat stronger than for
the $27\,M_\odot$ SN progenitor. Indeed, the detection rate for this
progenitor is slightly higher than for the $27\,M_\odot$ case.

Figure~\ref{fig:4pimap} (bottom panel) shows the sky-map of
$\sigma/\sigma_{\mathrm{max}}$ for the SASI episode 150--330~ms of the
$20\,M_\odot$ star. Comparing the two panels of Fig.~\ref{fig:4pimap},
we see that, as
already pointed out in Sec.~\ref{sec:2027Msun}, SASI develops for the
$20\,M_\odot$ progenitor on a different plane than for the $27\,M_\odot$
case. Therefore, the optimal observer directions to detect SASI effects are
almost perpendicular in the two models.

In our earlier paper~\cite{Tamborra:2013laa}, we have considered the power
spectrum of the IceCube rate for all the three studied SN progenitors.
One finds a strong peak at $f\simeq80$~Hz  for the $27$ and $20\,M_\odot$ cases, 
corresponding to the typical SASI
frequency. This frequency equals the one that describes large-amplitude
fluctuations of the low spherical harmonics SASI amplitude vector in  Fig.~2
(right panel on the top) of Ref.~\cite{Hanke:2013ena}. It is also
basically understood from analytic and numerical studies of the linear
growth regime of the SASI, and it is roughly the inverse of the
advection timescale plus the sound travel timescale (see Eq.~2 of 
Ref.~\cite{Tamborra:2013laa}). 
Both of these timescales only depend on shock radius and neutron-star
radius~\cite{Scheck:2007gw,Tamborra:2013laa}.

\section{Discussion and Summary}
\label{sec:discussion}

The first 3D full-scale hydrodynamical SN simulations with sophisticated
neutrino transport are now available for three SN progenitors with masses
11.2, 20 and $27\,M_\odot$, respectively. In a series of papers, we have
explored the neutrino emission properties of these models and in particular
the dependence on observer direction and the time variability of the signal
and opportunities to measure them in large-scale detectors such as IceCube
and the future Hyper-Kamiokande.

The first important point was made in our companion {\em Physical Review
Letter}~\cite{Tamborra:2013laa} where we emphasized the appearance of
pronounced SASI activity in our two heavier progenitors. The question if SASI
indeed appears in 3D models or if it would be suppressed by convective
overturn had been debated among SN modelers, but a consensus seems to be
appearing that SASI is not generically suppressed in 3D. Of course,
the appearance of SASI needs confirmation also by future 3D simulations
that yield successful explosions (none of our 3D models has led to an explosion
so far), and numerical
simulations might still be different from what happens in real stars.
Detecting SASI in the neutrino signal of the next nearby SN would go a long
way in testing our hydrodynamical understanding of stellar core collapse. With
IceCube and the future Hyper-Kamiokande, a galactic SN offers a realistic
opportunity for such a detection at any distance  up to 20~kpc, but the
signal amplitude strongly depends on the observer direction relative to the main
SASI plane of motion.

The main point of our present paper is to provide more details about the
neutrino signals of these models and their directional dependence. We stress
that observer-related quantities are weighted hemispheric averages with appropriate
flux-projection effects as considered here and
outlined in our Appendix~A. We have also provided,
in Appendix~B, simple analytic approximation formulas, based on the IBD cross
sections of Ref.~\cite{Strumia:2003zx}, that allow one to obtain detection rates
based on the parameters of an assumed Gamma distribution for the neutrino
spectra. In order to translate 3D model output into detection rates, SN
modelers would have to provide flavor-dependent luminosities as well as first and
second energy moments that are based on such observer-related hemispheric
averaging.

In our other companion paper~\cite{Tamborra:2014aua} we have reported a new
spherical-symmetry breaking effect in the form of LESA. The emission of
$\nu_e$ and $\bar\nu_e$, during the accretion phase, builds up a distinct
dipole pattern such that deleptonization happens predominantly in one
hemisphere. Therefore, the relative number fluxes of $\nu_e$ and $\bar\nu_e$
show a strong angular variation. It has not yet been explored what this means
in the context of flavor oscillations with neutrino-neutrino refractive
effects.

The direction of the LESA dipole and the plane of SASI sloshing and spiral
modes are apparently not related---these are different effects that can coexist. In 
particular, while we find LESA for all three studied progenitors, SASI occurs only for the 
heavier ones. Any influence of SASI on the LESA dipole orientation
seems to depend on 
the relative LESA and SASI dipole orientations, both randomly established for each progenitor.
LESA survives phases of violent SASI activity, even though it may be somewhat
masked by the latter. Further analysis on the LESA phenomenon and hydrodynamical simulations 
for more SN progenitors are needed to properly disentangle the two effects.

During the standing-shock accretion-powered phase of neutrino emission,
several new effects develop in 3D in contrast
to the traditional spherically-symmetric
picture. This phase offers a rich variety of new hydrodynamical and
neutrino-hydrodynamical phenomenology that has only begun to be explored. The
theory of neutrino flavor conversion with neutrino-neutrino refraction needs
to be further developed to understand its role during this phase. A future
high-statistics observation by IceCube and Hyper-Kamiokande will provide
opportunities to test such effects, and in particular the appearance of SASI
modes.

\subsection*{Acknowledgements}

We thank Ewald M\"uller for discussions.  This research was supported by the
Deutsche Forschungsgemeinschaft (DFG) through the Transregional Collaborative
Research Center SFB/TR 7 ``Gravitational Wave Astronomy'' and the Cluster of
Excellence EXC 153 ``Origin and Structure of the Universe''
(http://www.universe-cluster.de), and by the EU through ERC-AdG No.\
341157-COCO2CASA. I.T.\ acknowledges support from the
Netherlands Organization for Scientific Research (NWO). Our results could
only be achieved with the assistance of high performance computing resources
(Tier-0) provided by PRACE on CURIE TN (GENCI@CEA, France) and SuperMUC
(GCS@LRZ, Germany) and by the Gauss Centre for Supercomputing (GCS) on
SuperMUC. We also thank the Rechenzentrum Garching for computing
time on the IBM iDataPlex system \emph{hydra}.

\appendix

\section{Neutrino Flux Projections}
\label{sec:fluxprojections}

\begin{figure}[ht]
\centering
\includegraphics[width=0.9\columnwidth]{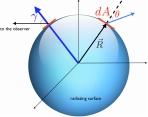}
\caption{Sketch of the quantities appearing in the calculation of the SN
neutrino flux measured by a distant observer. The neutrino  intensity is
defined in terms of the location $\mathbf{R}$ of the emitting surface element
$dA$ and the angle $\omega=(\theta,\phi)$ of emission
relative to the direction $\mathbf{R}$. The
observer is located at a distance $D  \gg R$ in an arbitrary direction
$\Omega=(\Theta,\Phi)$, and $\gamma$ is the angle between the location of the
radiating surface element and the direction of the observer.
\label{fig:projection}}
\end{figure}

Given the neutrino emission characteristics at the SN from a 3D simulation we
need to calculate the flux measurable by a distant observer, closely
following Ref.~\cite{Muller:2011yi}. Given a coordinate system in which the
simulation has been performed (see Fig.~\ref{fig:projection}), the observer
is located at a large distance $D\gg R$ in an arbitrary direction
$\Omega=(\Theta,\Phi)$. Here $R$ is the radius of a sphere near the SN
where the neutrino intensities are specified by the output of the code. We have
chosen $R=500~{\rm km}$ so that it is not necessary to apply
coordinate transformations and redshift effects between the fluid frame and
the distant observer. All quantities depend on time $t$, which we never show
explicitly, and we neglect retardation effects between neutrinos emitted from
different regions of the emitting surface.

We assume that the neutrino intensity $I({\bm R},{\omega})$ is given in terms
of the location ${\bm R}$ on the emitting surface. The angle
${\omega}=(\theta,\phi)$ describes the angular emission characteristic
relative to the direction ${\bm R}$ on the surface. While the intensity is
usually defined as the local spectral energy density of the neutrino
radiation field for a given direction of motion ${\omega}$ times the speed of
light, we here take it to be integrated over energy or over a specific energy
bin. It is trivial to go back to spectral quantities (differential with
regard to neutrino energy).

In order to obtain the energy flux at the location of the observer we have to
integrate over solid angles $d\Omega'$ over the surface of the source as seen
by the observer and add up the flux contributions emitted by each surface
element in the direction of the observer. A given area  $dA$ on the emitting
surface has the transverse cross section, as seen by the observer, of
$\cos\gamma\,dA$ where $\gamma$ is the angle between ${\bm R}$ (location of
the surface element) and the direction of the observer (see
Fig.~\ref{fig:projection}) so that $d\Omega'=\cos\gamma\,dA/D^2$. The
observable flux is therefore
\begin{equation}
F_\Omega=\frac{1}{D^2}\int\limits_{\rm{visible\atop surface}}dA\,
\cos\gamma\,I({\bm R},{\omega}_\Omega)\,,
\end{equation}
where $\omega_\Omega$ is the emission direction toward the observer and
$F_\Omega$ is the energy flux at distance $D$ in the direction $\Omega$. If
the observer interprets this flux as originating from a spherically symmetric
source, the measured flux corresponds to the $4\pi$-equivalent luminosity of
$4\pi D^2$ times this expression or
\begin{equation}
L_\Omega=4\pi\int\limits_{\rm{visible\atop surface}}dA\,
\cos\gamma\,I({\bm R},{\omega}_\Omega)\,,
\end{equation}
where the surface integral is over the part of the surface that is visible to
the observer.

Our 3D hydrodynamical simulations are based on the ray-by-ray scheme
\cite{Rampp:2002bq} where in each angular zone one solves a 1D neutrino
transport problem so that, within such a zone, the emission is axially
symmetric and depends only on the zenith angle, $\theta$, relative to the
radial direction. In principle, $I({\bm R},\theta)$ can be extracted from the
numerical results, but would require a vast amount of post-processing of huge
data files. Instead, we fall back on a simple approximation where the
directional distribution on each point of the radiating surface can be
described by the diffusion approximation for a radial
flux~\cite{Mihalas:1978,Muller:2011yi}
\begin{equation}
I({\bm R},\theta) = a\,E({\bm R}) + b\,F({\bm R}) \cos\theta\,,
\end{equation}
with $E$ the neutrino energy density and $F$ the neutrino energy flux. In
order to determine $a$ and $b$ we refer to the definitions of $E$
and $F$ in terms of angular integrals of the intensity $I$,
\begin{equation}
E({\bm R}) = \frac{2\pi}{c}\int_{-1}^{+1}I({\bm R},\theta)\,d\cos\theta
\end{equation}
and
\begin{equation}
F({\bm R})=2\pi \int_{-1}^{+1}I({\bm R},\theta)\,\cos\theta\;d\cos\theta\,,
\end{equation}
where $c$ is the speed of light. To express both coefficients by the
same quantity we assume $F= f\,c\,E$ so that $I({\bm
R},\theta)=(f^{-1}+3\cos\theta)\,F({\bm R})/4\pi$. The value of $f$ is
determined by the requirement that for $F=\mathrm{const.}$ on the sphere,
after integrating
over the entire surface, one obtains the luminosity $L= 4 \pi R^2 F$ so that
\begin{equation}
\label{decom}
I({\bm R},\theta) = \frac{F({\bm R})}{2\pi} \left(1+\frac{3}{2} \cos \theta\right)\,.
\end{equation}
Note that since $I\ge0$, Eq.~(\ref{decom}) is strictly valid only for $\cos
\theta \ge -2/3$, which includes inward going radiation for $\cos \theta <
0$. Equation~(\ref{decom}) reproduces the limb-darkening effect---see
Ref.~\cite{Mihalas:1978} for more details. With this result, we finally
obtain
\begin{equation}
L_\Omega=2 \int\limits_{\rm{visible\atop surface}} dA\,
\left(1+\frac{3}{2}\cos\gamma\right)\cos\gamma\; F({\bm R})\,,
\end{equation}
where we have inserted $\theta=\gamma$ because the zenith angle of local
radiation emission that points in the direction of the observer is identical
with $\gamma$ (see Fig.~\ref{fig:projection}). The value of $\gamma$ at a
surface point ${\bm R}$ depends, of course, on the observer direction
$\Omega$.

In conclusion, the only information from the numerical SN model that we actually use
is the radial neutrino-energy dependent flux $F({\bm R},\epsilon_\nu)$ on a given surface. From here,
we perform projections along the direction of the observer for the energy-dependent fluxes
before computing the observable $L$,
$\langle\epsilon_\nu\rangle$, $\langle\epsilon_\nu^2\rangle$
for the neutrino spectral information in the form of an assumed $\Gamma$ distribution
(see Appendix~\ref{sec:ibd}).

\section{Neutrino Spectra and Inverse Beta Cross Section}
\label{sec:ibd}

The quasi-thermal neutrino spectra produced at the SN can be well
approximated in terms of a Gamma distribution which has the normalized
form~\cite{Keil:2002in, Tamborra:2012ac},
\begin{equation}
f(\epsilon)=\frac{\epsilon^\alpha}{\Gamma_{\alpha+1}}\,\left(\frac{\alpha+1}{A}\right)^{\alpha+1}
\exp\left[-\frac{(\alpha+1)\,\epsilon}{A}\right]\,,
\end{equation}
where $\Gamma$ is the Gamma function, $A$ an energy scale, and
$\alpha$ a shape parameter with $\alpha=2$ corresponding to a
Maxwell-Boltzmann distribution. The spectra are usually ``pinched,''
meaning that usually $\alpha>2$. For the moments of the distribution
we use the notation
\begin{equation}
\epsilon_n=\langle\epsilon^n\rangle=\int_0^\infty d\epsilon\,\epsilon^n\,f(\epsilon)\,.
\end{equation}
The first two moments are
\begin{equation}
\epsilon_1=\langle \epsilon\rangle=A
\quad\hbox{and}\quad
\epsilon_2=\langle \epsilon^2\rangle=\frac{\alpha+2}{\alpha+1}\,A^2\,.
\end{equation}
This implies that the shape parameter is given in terms of the first
two moments as
\begin{equation}
\alpha=\frac{\epsilon_2-2\epsilon_1^2}{\epsilon_1^2-\epsilon_2}
=\frac{\langle\epsilon\rangle^2-\epsilon_{\rm rms}^2}{\epsilon_{\rm rms}^2}
\,.
\end{equation}
The rms width of the Gamma distribution is $\epsilon_{\rm rms} =
\sqrt{\langle \epsilon^2\rangle-\langle
\epsilon\rangle^2}=A/\sqrt{\alpha+1}$. From the numerical data we extract the
energy moments $\epsilon_1$ and $\epsilon_2$ and determine $A$ and $\alpha$
accordingly.

\begin{figure}[ht]
\centering
\vspace{5mm}\includegraphics[width=0.9\columnwidth]{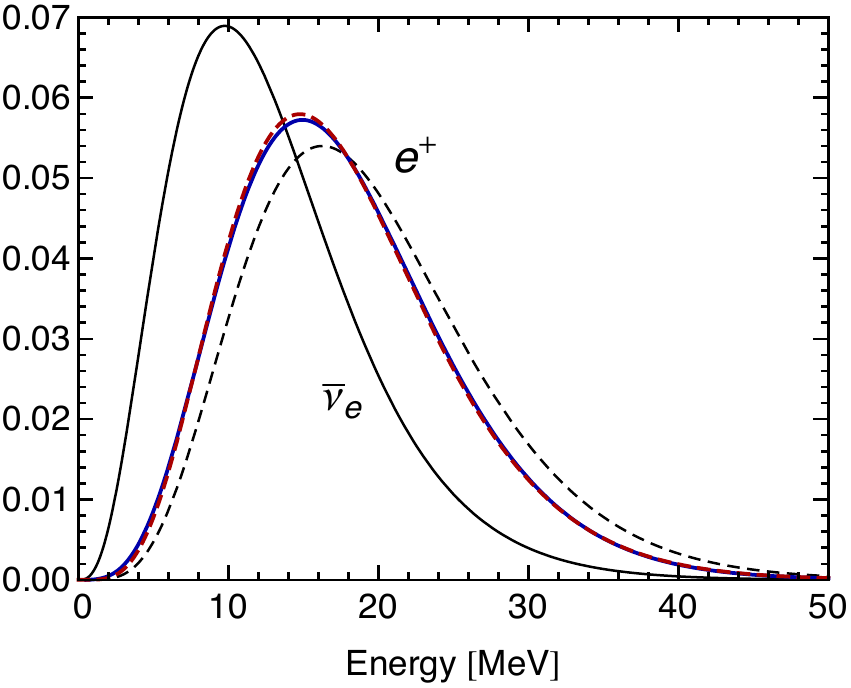}
\caption{Normalized energy distribution of $\bar\nu_e$, assuming a Gamma distribution
(solid black line) with $A_\nu=13$~MeV and $\alpha_\nu=3$. Corresponding normalized
positron distribution (solid blue line) after folding with inverse-beta decay
cross section. Approximation with a Gamma distribution (dashed red line) with
same $\langle\epsilon_e\rangle$ and $\langle\epsilon_e^2\rangle$. Assuming an
inverse-beta cross section
scaling with $\epsilon_\nu^2$ and no recoils gives the dashed black spectrum.
\label{fig:distribution}}
\end{figure}

The main detection process is inverse beta decay (IBD), $\bar\nu_e
p\to n e^+$, where the final-state positron shows up by its
Cherenkov radiation. Therefore, the primary $\bar\nu_e$ spectrum
must be translated to the corresponding $e^+$ spectrum. If we take
positrons to be massless, ignore the proton-neutron mass difference
as well as recoil effects, the cross section is
\begin{eqnarray}
\sigma_{\rm naive}&=&\frac{G_{\rm F}^2\cos^2\theta_{\rm C}\,(1+3C_A^2)}{\pi}\,
\epsilon_\nu^2\\
\nonumber
&=&9.343\times10^{-44}~{\rm cm}^2\,\left(\frac{\epsilon_\nu}{{\rm MeV}}\right)^2\,,
\end{eqnarray}
where $G_{\rm F}=1.166\times10^{-5}~{\rm GeV}^{-2}$ is the Fermi
constant, $\cos\theta_{\rm C}= 0.9746\pm 0.0008$ the cosine of the
Cabibbo angle, and $C_A=-1.270 \pm 0.003$ the axial-vector coupling
constant. With this simple $\epsilon_\nu^2$ scaling, the positron spectrum
would also follow a Gamma distribution with average energy
$A_e=A_\nu (3+\alpha_\nu)/(2+\alpha_\nu)$ and
$\alpha_e=\alpha_\nu+2$. An example with $A_\nu=13$~MeV and
$\alpha_\nu=3$ is shown in Fig.~\ref{fig:distribution} as a thin
solid line. The corresponding naive positron distribution has
$A_e=\frac{3}{2}\,A_\nu=19.5$~MeV and $\alpha_e=5$, shown as a thin
dashed black line.

A realistic IBD cross section requires to take into account recoil effects,
the neutron-proton mass difference, the positron mass, and nucleon form
factors. We use the results of Ref.~\cite{Strumia:2003zx} to
derive the positron distribution, shown as a thick blue line in
Fig.~\ref{fig:distribution}. We compare it with a Gamma distribution (dashed
red line) with the same $\langle\epsilon\rangle$ and
$\langle\epsilon^2\rangle$ and find $A_e=17.86$~MeV and $\alpha_e=4.76$. We
show a normalized spectrum here; the average cross section is approximately
0.74 of the naive result. We conclude that the positron spectrum is also well
approximated by a Gamma distribution. (Of course, the positron spectrum
strictly begins only at $\epsilon_e=m_e$, but the energy range below a few
MeV is irrelevant in practice.)

What remains is to express the positron Gamma-distribution
parameters in terms of those of the primary $\bar\nu_e$ spectrum. We
have derived analytic approximation functions for this
transformation. Expressing all energies in MeV, we find
\begin{eqnarray}
A_e&=&\sqrt{m_e^2+\left(\frac{3+\alpha_\nu}{1+\alpha_\nu}\,A_\nu\right)^2}
-\frac{5A_\nu}{4(8+\alpha_\nu)}
\nonumber\\*
&\times&\left(\frac{31+9\,\alpha_\nu}{10+7 A_\nu}+
\frac{170+47\,\alpha_\nu+\alpha_\nu^2}{1+4\alpha_\nu}\,\frac{A_\nu}{600}\right)\,,
\\[2ex]
\alpha_e&=&2.25+\frac{0.4}{1+(A_\nu/3.7)^{3.5}}
+\frac{1.08\,A_\nu-0.7}{3+A_\nu}\,\alpha_\nu\,, \\[2ex]
\langle\sigma\rangle&=&7.37\times10^{-46}~{\rm cm}^2~
\frac{2+\alpha_\nu}{1+\alpha_\nu}\,A_\nu^{2.15}\\
\nonumber\\*
&\times&\left(\frac{76.64}{\alpha_\nu^{0.021}}-\frac{A_\nu}{\alpha_\nu^{0.24}}\right)
\nonumber\\*
&\times&\left[1-\exp\left(\frac{-0.25 + 0.55\,\alpha_\nu}{2.2+\alpha_\nu}
-\frac{1+1.6\,\alpha_\nu}{1+4\,\alpha_\nu}\,A_\nu\right)\right]\,.
\nonumber
\end{eqnarray}
Typically, these approximation formulas are good to much better than
1\% in our range of interest.



\begin{thebibliography}{00}

\bibitem{Scholberg:2012id}
  K.~Scholberg,
  Ann.\ Rev.\ Nucl.\ Part.\ Sci.\  {\bf 62}, 81 (2012).

\bibitem{Abe:2013gga}
  K.~Abe {\it et al.} (Super-Kamiokande Collaboration),
  Nucl.\ Instrum.\ Meth.\ A {\bf 737}, 253 (2014).

\bibitem{Mori:2013wua}
  T.~Mori (for Super-Kamiokande Collaboration),
  Nucl.\ Instrum.\ Meth.\ A {\bf 732}, 316 (2013).

\bibitem{Abbasi:2011ss}
  R.~Abbasi {\it et al.} (IceCube Collaboration),
  Astron.\ Astrophys.\  {\bf 535}, A109 (2011).

\bibitem{Demiroers:2011am}
  M.~Salathe, M.~Ribordy and L.~Demirors,
  Astropart.\ Phys.\ {\bf 35}, 485 (2012).

\bibitem{Aartsen:2013nla}
  M.~G.~Aartsen {\it et al.} (IceCube Collaboration),
  arXiv: 1309.7008.

\bibitem{Li:2014qca}
  Y.-F.~Li,
  Int.\ J.\ Mod.\ Phys.\ Conf.\ Ser.\  {\bf 31}, 1460300 (2014).

\bibitem{Abe:2011ts}
  K.~Abe {\it et al.} (Hyper-Kamiokande Working Group),
  arXiv:1109.3262.

\bibitem{Agarwalla:2013kaa}
  S.~K.~Agarwalla {\it et al.} (LAGUNA-LBNO Collaboration),
  JHEP {\bf 1405}, 094 (2014).

\bibitem{LBNE}
  Long-Baseline Neutrino Experiment Document Data\-base,
 {\tt http://lbne2-docdb.fnal.gov/}

\bibitem{vandenBergh:1994}
  S.~van den Bergh and R.~D.~McClure,
  Astrophys.\ J.\ {\bf 425}, 205 (1994).

\bibitem{Li:2010kd}
  W.~Li {\it et al.},
  Mon.\ Not.\ R.\ Astron.\ Soc.\ {\bf 412}, 1473 (2011).


\bibitem{Reed:2005en}
  B.~C.~Reed,
  Astron.\ J.\  {\bf 130}, 1652 (2005).


\bibitem{FaucherGiguere:2005ny}
  C.-A.~Faucher-Gigu\`ere and V.~M.~Kaspi,
  Astrophys.\ J.\  {\bf 643}, 332 (2006).


\bibitem{Keane:2008jj}
  E.~F.~Keane and M.~Kramer,
  Mon.\ Not.\ R.\ Astron.\ Soc.\ {\bf 391}, 2009 (2008).


\bibitem{Diehl:2006cf}
  R.~Diehl {\it et al.},
  Nature {\bf 439}, 45 (2006).

\bibitem{Strom:1994}
  R.~G.~Strom,
  Astron.\ Astrophys.\ {\bf 288}, L1 (1994).

\bibitem{Tammann:1994ev}
  G.~A.~Tammann, W.~L\"offler and A.~Schr\"oder,
  Astrophys.\ J.\ Suppl.\  {\bf 92}, 487 (1994).

\bibitem{Adams:2013ana}
  S.~M.~Adams, C.~S.~Kochanek, J.~F.~Beacom, M.~R.~Vagins and K.~Z.~Stanek,
  Astrophys.\ J.\  {\bf 778}, 164 (2013).


\bibitem{Alekseev:1993dy}
  E.~N.~Alekseev {\it et al.},
  Zh.\ Eksp.\ Teor.\ Fiz.\  {\bf 104}, 2897 (1993)
  [J.\ Exp.\ Theor.\ Phys.\ {\bf 77}, 339 (1993)].

\bibitem{Schramm:1987ra}
  D.~N.~Schramm,
  Comments Nucl.\ Part.\ Phys.\  {\bf 17}, 239 (1987).

\bibitem{Raffelt:1990yz}
  G.~G.~Raffelt,
  Phys.\ Rept.\  {\bf 198}, 1 (1990).

\bibitem{Raffelt:1999tx}
  G.~G.~Raffelt,
  Ann.\ Rev.\ Nucl.\ Part.\ Sci.\  {\bf 49}, 163 (1999).

\bibitem{Totani:1997vj}
  T.~Totani, K.~Sato, H.~E.~Dalhed and J.~R.~Wilson,
  Astrophys.\ J.\  {\bf 496}, 216 (1998).


\bibitem{Lund:2010kh}
  T.~Lund, A.~Marek, C.~Lunardini, H.-T.~Janka and G.~G.~Raffelt,
  Phys.\ Rev.\ D {\bf 82}, 063007 (2010).

\bibitem{Lund:2012vm}
  T.~Lund, A.~Wongwathanarat, H.-T.~Janka, E.~M\"uller and G.~G.~Raffelt,
  Phys.\ Rev.\ D {\bf 86}, 105031 (2012).

\bibitem{Brandt:2010xa}
  T.~D.~Brandt, A.~Burrows, C.~D.~Ott and E.~Livne,
  Astrophys.\ J.\  {\bf 728}, 8 (2011).

\bibitem{Tamborra:2013laa}
  I.~Tamborra, F.~Hanke, B.~M\"uller, H.-T.~Janka and G.~G.~Raffelt,
  Phys.\  Rev.\  Lett.\  {\bf 111}, 121104 (2013).

\bibitem{OConnor:2013} 
E. O'Connor, C.~D.~Ott,
Astrophys.\ J.\  {\bf 762}, 126 (2013).


\bibitem{Mueller:2014rna}
  B.~M\"uller and H.-T.~Janka,
Astrophys.\ J.\  {\bf 788}, 82 (2014).


\bibitem{Bethe:1984ux}
  H.~A.~Bethe and J.~R.~Wilson,
  Astrophys.\ J.\  {\bf 295}, 14 (1985).

\bibitem{Bethe:1990mw}
  H.~A.~Bethe,
  Rev.\ Mod.\ Phys.\  {\bf 62}, 801 (1990).

\bibitem{Janka:2012wk}
  H.-T.~Janka,
  Ann.\ Rev.\ Nucl.\ Part.\ Sci.\  {\bf 62}, 407 (2012).


\bibitem{Herant:1994dd}
  M.~Herant, W.~Benz, W.~R.~Hix, C.~L.~Fryer and S.~A.~Colgate,
  Astrophys.\ J.\  {\bf 435}, 339 (1994).


\bibitem{Burrows:1995ww}
  A.~Burrows, J.~Hayes and B.~A.~Fryxell,
  Astrophys.\ J.\  {\bf 450}, 830 (1995).


\bibitem{Janka:1996}
H.-T.~Janka and E.~M\"uller,
Astron.\ Astrophys.\ {\bf 306}, 167 (1996).

\bibitem{Marek:2008qi}
  A.~Marek, H.-T.~Janka and E.~M\"uller,
  Astron.\ Astrophys.\ {\bf 496}, 475 (2009).


\bibitem{Marek:2007gr}
  A.~Marek and H.-T.~Janka,
  Astrophys.\ J.\  {\bf 694}, 664 (2009).

\bibitem{Mueller:2012is}
  B.~M\"uller, H.-T.~Janka and A.~Marek,
  Astrophys.\ J.\  {\bf 756}, 84 (2012).

\bibitem{Mueller:2012ak}
  B.~M\"uller, H.-T.~Janka and A.~Heger,
  Astrophys.\ J.\  {\bf 761}, 72 (2012).

\bibitem{Murphy:2008dw}
  J.~W.~Murphy and A.~Burrows,
  Astrophys.\ J.\  {\bf 688}, 1159 (2008).

\bibitem{Nordhaus:2010uk}
  J.~Nordhaus, A.~Burrows, A.~Almgren and J.~Bell,
  Astrophys.\ J.\  {\bf 720}, 694 (2010).

\bibitem{Fryer:2002}
C.~L.~Fryer and M.~S.~Warren,
Astrophys.\ J.\  {\bf 574}, L65 (2002).

\bibitem{Fryer:2003jj}
  C.~L.~Fryer and M.~S.~Warren,
Astrophys.\ J.\  {\bf 601}, 391 (2004).


\bibitem{Iwakami:2007ie}
  W.~Iwakami, K.~Kotake, N.~Ohnishi, S.~Yamada and K.~Sawada,
  Astrophys.\ J.\  {\bf 678}, 1207 (2008).

\bibitem{Wongwathanarat:2010ip}
  A.~Wongwathanarat, H.-T.~Janka and E.~M\"uller,
  Astrophys.\ J.\  {\bf 725}, L106 (2010).

\bibitem{Hanke:2011jf}
  F.~Hanke, A.~Marek, B.~M\"uller and H.-T.~Janka,
  Astrophys.\ J.\  {\bf 755}, 138 (2012).

\bibitem{Takiwaki:2011db}
  T.~Takiwaki, K.~Kotake and Y.~Suwa,
  Astrophys.\ J.\  {\bf 749}, 98 (2012).

\bibitem{Burrows:2012yk}
  A.~Burrows, J.~C.~Dolence and J.~W.~Murphy,
  Astrophys.\ J.\  {\bf 759}, 5 (2012).

\bibitem{Ott:2013}
  C.~D.~Ott {\it et al.},
  Astrophys.\ J.\  {\bf 768}, 115 (2013).

\bibitem{Muller:2011yi}
  E.~M\"uller, H.-T.~Janka and A.~Wongwathanarat,
  Astron.\ Astrophys.\ {\bf 537}, 63 (2012).

\bibitem{Blondin:2002sm}
  J.~M.~Blondin, A.~Mezzacappa and C.~DeMarino,
  Astrophys.\ J.\  {\bf 584}, 971 (2003).

\bibitem{Scheck:2007gw}
  L.~Scheck, H.-T.~Janka, T.~Foglizzo and K.~Kifonidis,
  Astron.\ Astrophys.\  {\bf 477}, 931 (2008).

\bibitem{Foglizzo:2011aa}
  T.~Foglizzo, F.~Masset, J.~Guilet and G.~Durand,
  Phys.\ Rev.\ Lett.\  {\bf 108}, 051103 (2012).

\bibitem{Tamborra:2014aua}
  I.~Tamborra, F.~Hanke, H.-T.~Janka, B.~M\"uller, G.~G.\ Raffelt and A.~Marek,
   Astrophys.\ J.\  {\bf 792}, 96 (2014).

\bibitem{Murphy:2012id}
  J.~W.~Murphy, J.~C.~Dolence and A.~Burrows,
  Astrophys.\ J.\  {\bf 771}, 52 (2013).

\bibitem{Dolence:2012kh}
  J.~C.~Dolence, A.~Burrows, J.~W.~Murphy and J.~Nordhaus,
  Astrophys.\ J.\  {\bf 765}, 110 (2013).



\bibitem{Hanke:2013ena}
  F.~Hanke, B.~M\"uller, A.~Wongwathanarat, A.~Marek and H.-T.~Janka,
  Astrophys.\ J.\  {\bf 770}, 66 (2013).

\bibitem{Couch:2013kma}
  S.~M.~Couch and E.~P.~O'Connor,
 Astrophys.\ J.\  {\bf 785}, 123 (2014).

\bibitem{Sumiyoshi:2014qua}
  K.~Sumiyoshi, T.~Takiwaki, H.~Matsufuru and S.~Yamada,
  arXiv:1403.4476.

\bibitem{Dolence:2014rwa}
  J.~C.~Dolence, A.~Burrows and W.~Zhang,
  arXiv: 1403.6115.


\bibitem{Kachelriess:2004ds}
  M.~Kachelriess, R.~Tom\`as, R.~Buras, H.-T.~Janka, A.~Marek and M.~Rampp,
  Phys.\ Rev.\ D {\bf 71}, 063003 (2005).

\bibitem{Serpico:2011ir}
  P.~D.~Serpico, S.~Chakraborty, T.~Fischer, L.~H\"udepohl, H.-T.~Janka and A.~Mirizzi,
  Phys.\ Rev.\ D {\bf 85}, 085031 (2012).

\bibitem{Rampp:2002bq}
  M.~Rampp and H.-T.~Janka,
  Astron.\ Astrophys.\  {\bf 396}, 361 (2002).

\bibitem{Buras:2005rp}
  R.~Buras, M.~Rampp, H.-T.~Janka and K.~Kifonidis,
  Astron.\ Astrophys.\  {\bf 447}, 1049 (2006).

\bibitem{Marek:2005if}
  A.~Marek, H.~Dimmelmeier, H.-T.~Janka, E.~M\"uller and R.~Buras,
  Astron.\ Astrophys.\  {\bf 445}, 273 (2006).

\bibitem{Woosley:2002}
  S.~E.~Woosley, A.~Heger and T.~A.~Weaver,
  Rev.\ Mod.\ Phys.\ {\bf 74}, 1015 (2002).

\bibitem{Woosley:2007}
  S.~E.~Woosley and A.~Heger,
  Phys.\ Rep.\ {\bf 442}, 269 (2007).

\bibitem{Lattimer:1991nc}
  J.~M.~Lattimer and F.~D.~Swesty,
  Nucl.\ Phys.\ A {\bf 535}, 331 (1991).

\bibitem{Bruenn:2012mj}
  S.~W.~Bruenn {\it et al.},
  Astrophys.\ J.\  {\bf 767}, L6 (2013).


\bibitem{Keil:2002in}
  M.~T.~Keil, G.~G.~Raffelt and H.-T.~Janka,
  Astrophys.\ J.\  {\bf 590}, 971 (2003).

\bibitem{Wolfenstein:1977ue}
  L.~Wolfenstein,
  Phys.\ Rev.\ D {\bf 17}, 2369 (1978).

  \bibitem{wolf}
  S.~P.~Mikheev and A.~Yu.~Smirnov,
  Yad. Fiz. \textbf{42}, 1441 (1985)  [Sov. J. Nucl. Phys. \textbf{42}, 913 (1985)].

\bibitem{Beringer:1900zz}
  J.~Beringer {\it et al.} (Particle Data Group Collaboration),
  Phys.\ Rev.\ D {\bf 86}, 010001 (2012).

\bibitem{Dighe:1999bi}
  A.~S.~Dighe and A.~Yu.~Smirnov,
  Phys.\ Rev.\ D {\bf 62}, 033007 (2000).

\bibitem{Kifonidis:2003fv}
  K.~Kifonidis, T.~Plewa, H.-T.~Janka and E.~M\"uller,
ÊÊAstron.\ Astrophys.\  {\bf 408}, 621 (2003).
Ê


\bibitem{Kifonidis:2005yj}
  K.~Kifonidis, T.~Plewa, L.~Scheck, H.-T.~Janka and E.~M\"ueller,
ÊAstron.\ Astrophys.\  {\bf 453}, 661 (2006).


\bibitem{Scheck:2006rw}
  L.~Scheck, K.~Kifonidis, H.-T.~Janka and E.~M\"ueller,
ÊAstron.\ Astrophys.\  {\bf 457}, 963 (2006).


\bibitem{Hammer:2009cn}
  N.~J.~Hammer, H.-T.~Janka and E.~M\"ueller,
ÊÊAstrophys.\ J.\  {\bf 714}, 1371 (2010).
Ê


\bibitem{Loreti:1995ae}
  F.~N.~Loreti, Y.~Z.~Qian, G.~M.~Fuller and A.~B.~Balantekin,
ÊÊPhys.\ Rev.\ D {\bf 52}, 6664 (1995).
Ê


\bibitem{Fogli:2006xy}
  G.~L.~Fogli, E.~Lisi, A.~Mirizzi and D.~Montanino,
ÊÊJCAP {\bf 0606}, 012 (2006).
Ê


\bibitem{Friedland:2006ta}
  A.~Friedland and A.~Gruzinov,
ÊÊastro-ph/0607244.

\bibitem{Kneller:2010sc}
  J.~P.~Kneller and C.~Volpe,
ÊÊPhys.\ Rev.\ D {\bf 82} 123004, (2010).
Ê


\bibitem{Lund:2013uta}
  T.~Lund and J.~P.~Kneller,
ÊÊPhys.\ Rev.\ D {\bf 88}, 2,  023008 (2013).
Ê

\bibitem{Borriello:2013tha}
  E.~Borriello, S.~Chakraborty, H.-T~Janka, E.~Lisi and A.~Mirizzi,
ÊÊarXiv:1310.7488 [astro-ph.SR].


\bibitem{Duan:2010bg}
  H.~Duan, G.~M.~Fuller and Y.-Z.~Qian,
  Ann.\ Rev.\ Nucl.\ Part.\ Sci.\  {\bf 60}, 569 (2010).

\bibitem{Duan:2006an}
  H.~Duan, G.~M.~Fuller, J.~Carlson and Y.-Z.~Qian,
  Phys.\ Rev.\ D {\bf 74}, 105014 (2006).

\bibitem{Fogli:2007bk}
  G.~L.~Fogli, E.~Lisi, A.~Marrone and A.~Mirizzi,
  JCAP {\bf 0712}, 010 (2007).


\bibitem{Raffelt:2007cb}
  G.~G.~Raffelt and A.~Yu.~Smirnov,
  Phys.\ Rev.\ D {\bf 76}, 081301 (2007);
  Erratum {\it ibid.} {\bf 77}, 029903 (2008).

\bibitem{Fogli:2008pt}
  G.~L.~Fogli, E.~Lisi, A.~Marrone, A.~Mirizzi and I.~Tamborra,
  Phys.\ Rev.\ D {\bf 78}, 097301 (2008).

\bibitem{Dasgupta:2009mg}
  B.~Dasgupta, A.~Dighe, G.~G.~Raffelt and A.~Yu.~Smirnov,
  Phys.\ Rev.\ Lett.\  {\bf 103}, 051105 (2009).

\bibitem{Fogli:2009rd}
  G.~L.~Fogli, E.~Lisi, A.~Marrone and I.~Tamborra,
ÊÊJCAP {\bf 0910}, 002 (2009).
Ê

\bibitem{Dasgupta:2010cd}
  B.~Dasgupta, A.~Mirizzi, I.~Tamborra and R.~Tom\`as,
ÊÊPhys.\ Rev.\ D {\bf 81}, 093008 (2010).
Ê


\bibitem{EstebanPretel:2008ni}
  A.~Esteban-Pretel {\em et al.},
  Phys.\ Rev.\ D {\bf 78}, 085012 (2008).

\bibitem{Chakraborty:2011nf}
  S.~Chakraborty, T.~Fischer, A.~Mirizzi, N.~Saviano and R.~Tom\`as,
  Phys.\ Rev.\ Lett.\  {\bf 107}, 151101 (2011);
%
  Phys.\ Rev.\ D {\bf 84},  025002 (2011).

\bibitem{Sarikas:2011am}
  S.~Sarikas, G.~G.~Raffelt, L.~H\"udepohl and H.-T.~Janka,
  Phys.\ Rev.\ Lett.\  {\bf 108}, 061101 (2012).

\bibitem{Saviano:2012yh}
  N.~Saviano, S.~Chakraborty, T.~Fischer and A.~Mirizzi,
  Phys.\ Rev.\ D {\bf 85}, 113002 (2012).

\bibitem{Cherry:2012zw}
  J.~F.~Cherry, J.~Carlson, A.~Friedland, G.~M.~Fuller and A.~Vlasenko,
  Phys.\ Rev.\ Lett.\  {\bf 108}, 261104 (2012).

\bibitem{Sarikas:2012vb}
  S.~Sarikas, I.~Tamborra, G.~Raffelt, L.~H\"udepohl and H.-T.~Janka,
  Phys.\ Rev.\ D {\bf 85}, 113007 (2012).

\bibitem{Raffelt:2013rqa}
  G.~Raffelt, S.~Sarikas and D.~de~Sousa~Seixas,
  Phys.\ Rev.\ Lett.\  {\bf 111}, 091101 (2013).

\bibitem{Raffelt:2013isa}
  G.~Raffelt and D.~de~Sousa~Seixas,
  Phys.\ Rev.\ D {\bf 88}, 045031 (2013).
 

\bibitem{Hansen:2014paa}
  R.~S.~Hansen and S.~Hannestad,
  Phys.\ Rev.\ D {\bf 90}, 025009 (2014).

\bibitem{Mirizzi:2013rla}
  A.~Mirizzi,
  Phys.\ Rev.\ D {\bf 88}, 073004 (2013).


\bibitem{Mirizzi:2013wda}
  S.~Chakraborty and A.~Mirizzi,
   Phys.\ Rev.\ D {\bf 90}, 033004 (2014).
  

\bibitem{Chakraborty:2014nma}
  S.~Chakraborty, A.~Mirizzi, N.~Saviano and D.~de~Sousa~Seixas,
  Phys.\ Rev.\ D {\bf 89}, 093001 (2014).
  
\bibitem{Mangano:2014zda}
  G.~Mangano, A.~Mirizzi and N.~Saviano,
  Phys.\ Rev.\ D {\bf 89}, 073017 (2014).

\bibitem{Strumia:2003zx}
  A.~Strumia and F.~Vissani,
  Phys.\ Lett.\ B {\bf 564}, 42 (2003).

\bibitem{Mihalas:1978}
  D.~Mihalas,
  {\em Stellar atmospheres\/}
  (W.~H.~Freeman, San Francisco, 1978).


\bibitem{Tamborra:2012ac}
  I.~Tamborra, B.~M\"uller, L.~H\"udepohl, H.-T.~Janka and G.~Raffelt,
  Phys.\ Rev.\ D {\bf 86}, 125031 (2012).

\end{thebibliography}
\end{document}